\documentclass[letterpaper,twocolumn,10pt]{article}
\usepackage{usenix,epsfig,endnotes}
\usepackage{hyperref}
\usepackage{ulem,bookmark}
\usepackage{url}
\usepackage{color}
\usepackage{tabularray}
\usepackage[utf8]{inputenc}
\usepackage{multirow} 
\usepackage{threeparttable}  
\usepackage{pifont}
\usepackage{bbding}
\usepackage{fontawesome}
\usepackage{tabularx}
\usepackage{cite}
\usepackage{amsmath,amssymb,amsfonts}
\usepackage{graphicx}
\usepackage{amsthm}%
\usepackage{mathrsfs}%
\usepackage[title]{appendix}%
\usepackage{manyfoot}%
\usepackage{algorithm}%
\usepackage{algpseudocode}%
\usepackage{listings}%
\usepackage{siunitx}
\usepackage{indentfirst}
\usepackage{caption}
\usepackage{subfigure}
\usepackage{subcaption}
\usepackage{xspace} 
\usepackage{soul}
\usepackage{titlesec}
\usepackage{makecell}
\usepackage{array}
\usepackage{framed}
\usepackage{booktabs}
\usepackage{pdfpages}
\usepackage{microtype}

\microtypecontext{spacing=nonfrench}
\usepackage{authblk}

\begin{document}
\pagestyle{empty}
\newcommand\gr[1]{{\color{blue}{\textbf{\{gr: {\em#1}\}}}}}
\newcommand\lzy[1]{{\color{black}{{#1}}}}
\newcommand\lzw[1]{{\color{purple}{\textbf{\{lzw: {\em#1}\}}}}}
\newcommand\fix[1]{{\color{red}{\textbf{\{fix: {\em#1}\}}}}}
\newcommand\modi[1]{{\color{red}{\textbf{\{modi: {\em#1}\}}}}}
\newcommand\zmm[1]{{\color{purple}{\textbf{\{zmm: {\em#1}\}}}}}
\newcommand{\toolname}{DVAHunter\xspace}
\newcommand{\domaindeploying}{domain deploying\xspace}
\newcommand{\requestforwarding}{request forwarding\xspace}

\hbadness=99999

\titlespacing{\section}{0pt}{2ex plus 1ex minus .2ex}{1ex plus .2ex}
\titlespacing{\subsection}{0pt}{1.5ex plus .2ex}{1ex plus .2ex}

\date{}

\title{\Large \bf Detecting and Measuring Security Implications \\ of Entangled Domain Verification in CDN}

\author[$\dagger\ddagger\rVert$]{Ziyu Lin}
\author[$\ast\ddagger$]{Zhiwei Lin}
\author[$\ddagger$]{Run Guo}
\author[$\ddagger$]{Jianjun Chen\textsuperscript{\ding{41}}}
\author[$\wr$]{Mingming Zhang}
\author[$\dagger$]{Ximeng Liu\textsuperscript{\ding{41}}}
\author[$\ddagger$]{Tianhao Yang}
\author[$\star$]{Zhuoran Cao}
\author[$\rVert$]{Robert H. Deng}

\affil[$\dagger$]{Fuzhou University}
\affil[$\ddagger$]{Tsinghua University}
\affil[$\ast$]{National University of Singapore}
\affil[$\wr$]{Zhongguancun Laboratory}
\affil[$\rVert$]{Singapore Management University}
\affil[$\star$]{Nanjing University of Posts and Telecommunications}

\maketitle

\footnotetext[1]{Corresponding authors: snbnix@gmail.com, jianjun@tsinghua.edu.cn}

\thispagestyle{empty}

\begin{abstract}
Content Delivery Networks (CDNs) offer a protection layer for enhancing the security of websites.
However, a significant security flaw named Absence of Domain Verification (DVA) has become emerging recently.
Although this threat is recognized, the current practices and security flaws of domain verification strategies in CDNs have not been thoroughly investigated.
In this paper, we present \toolname, an automated system for detecting DVA vulnerabilities that can lead to domain abuse in CDNs.
Our evaluation of 45 major CDN providers reveals the prevalence of DVA: most (39/45) providers do not perform any verification, and even those that do remain exploitable.
Additionally, we used \toolname to conduct a large-scale measurement of 89M subdomains from Tranco's Top 1M sites hosted on the 45 CDNs under evaluation. 
Our focus was on two primary DVA exploitation scenarios: covert communication and domain hijacking.
We identified over 332K subdomains vulnerable to domain abuse.
This tool provides deeper insights into DVA exploitation and allows us to propose viable mitigation practices for CDN providers. 
To date, we have received vulnerability confirmations from 12 providers; 6 (e.g., Edgio, Kuocai) have implemented fixes, and 1 (ChinaNetCenter) are actively working on solutions based on our recommendations.
\end{abstract}

\section{Introduction}

Content Delivery Network (CDN) is a critical Internet infrastructure designed to improve website security and reliability. 
Reports show that over 47.5\% of websites have delegated their major business domain names to CDN platforms~\cite{CDN_Usage_Statistics}. 
However, vulnerabilities in public CDNs could lead to widespread abuse of hosted domain names. 
In recent years, many prominent domain names hosted on CDNs have been abused by cyber attackers, including subdomains of Microsoft~\cite{Microsoft_takover} and major media like CCTV~\cite{CCTV}, which has brought about huge impacts.

Previous research has highlighted that the \textit{Domain Verification Absence} (\textit{DVA}) during \domaindeploying and \requestforwarding facilitates domain abuse in the CDN environment~\cite{measuring_domain_froting, domain_borrowing_bh, dare_shark}.
Attackers could exploit DVA to abuse domain names without authority on CDN platforms for malicious purposes like distributing malware and evading network censorship~\cite{cloud_strife,understand_dns_threat,google_hijacked,wall_street}.
Depending on the DNS configuration of victim domains, including the presence of dangling records in their nameservers, such domain abuse typically manifests as covert communication (e.g., domain fronting~\cite{domain_fronting} and domain borrowing~\cite{domain_borrowing_bh}) and domain hijacking (e.g., domain takeover~\cite{understand_dns_threat}).
However, existing work has not provided a comprehensive view of domain verification throughout the entire lifecycle of \domaindeploying and \requestforwarding in CDNs. 
Additionally, there is a need for an Internet-scale assessment of current state-of-the-art domain verification mechanisms and the prevalence of domain abuse threats in CDNs.

\noindent \textbf{Our study.} 
In this paper, we aim to systematically investigate domain verification mechanisms and their security flaws throughout the \textit{\domaindeploying} and \textit{\requestforwarding} phases within CDN environments.
We introduce \toolname, a comprehensive automated detection system designed to identify DVA vulnerabilities across mainstream CDN providers.  
We apply this system to conduct a large-scale measurement study aimed at identifying domain names potentially vulnerable to abuse threats among 89,015,946 subdomains under Tranco Top 1M domain names~\cite{tranco}.

Our measurement results show that 337,284 of the collected subdomains are hosted by 45 CDN providers. 
Among them, we discover that 43 providers, including Cloudflare~\cite{takeover_cloudflare}, Fastly~\cite{takeover_fastly}, and Netlify~\cite{takeover_netlify}, exhibit DVA vulnerabilities, exposing 332,710 (98.6\%) subdomains to domain abuse threats.
To gain new insights and assist in threat mitigation, we conduct an in-depth evaluation of the domain verification policies adopted by mainstream CDN providers. 
The evaluation reveals that while some CDN providers have deployed domain verification as a defense mechanism, we identified new flawed implementations that attackers can exploit to bypass the domain verification.
Our research has also identified a previously unknown security vulnerability associated with Multi-CDN platforms. 
These platforms utilize shared infrastructure across collaborating CDNs to extend global reach and enhance content delivery efficiency~\cite{multi-cdn}.
However, we demonstrate a critical risk that attackers can exploit Multi-CDNs (e.g., AligeCDN) for covert communication through domain fronting, hiding malicious traffic under benign domains of CDN providers like CloudFront.
The attackers could also leverage Multi-CDNs (e.g., KuaiKuaiCloud) to bypass the domain verification implemented by a secure CDN (e.g., Baidu) among the collaborators, highlighting that the overall security posture is compromised by the CDN with the least robust domain verification.

\noindent \textbf{Contributions.} We make the following contributions in this paper.
\begin{itemize}

\item[$\bullet$] \textit{New Attack Surface.}
We identified new flawed implementations that attackers can exploit to bypass the domain verification and take over domains. Our research has also identified a previously unknown security vulnerability associated with Multi-CDN platforms. 

\item[$\bullet$] \textit{A novel detection system.}
We propose \toolname, a novel system that automatically and periodically monitors DVA vulnerability at an Internet scale. Unlike previous studies focused on individual vulnerabilities, \toolname performs a comprehensive analysis of CDN domain verification and systematically measures vulnerabilities on a large scale.

\item[$\bullet$] \textit{Large-scale measurements.}
We use \toolname to evaluate 89,015,946 subdomains from Tranco Top 1M domains, affecting 332,710 subdomains hosted by CDN providers and discovering 43 CDN providers vulnerable to DVA, 20,296 borrowing domains, and 1449 dangling domains in vulnerable CDNs.

\item[$\bullet$] \textit{Publish CDN-related characteristics and open-source \toolname.} 
Firstly, we will publish the holistic characteristics of 45 CDN providers. Secondly, we will release \toolname through GitHub for researchers to further study DVA vulnerability in the future and help domain owners monitor their domain status.

\end{itemize}

\section{Background and Attack Surface Analysis}

\subsection{Content Delivery Networks}

CDN is an essential internet infrastructure consisting of a globally distributed cluster of servers. It mitigates website load by caching resources and accelerates user access time by deploying massive CDN nodes worldwide.
With the development of CDNs, to further improve performance, Multi-CDN has been proposed~\cite{multi-cdn, characterizing_multi-cdn}. 
Multi-CDN involves integrating multiple CDNs from different providers into a single network.
By adopting Multi-CDN, you can access a vast network of nodes from multiple CDNs, which can significantly enhance the speed of content delivery, expand web services regional and global coverage, and mitigate cybersecurity risks. 

From the user's perspective, when they try to access a website hosted on a CDN, the CDN's request routing mechanism~\cite{rfc3568} redirects the HTTP request to the ingress node that is best suited for that request. For example, in the ``DNS-based routing'' mechanism, the website domain is first resolved to a subdomain assigned by the CDN. Then, the CDN's DNS system~\cite{fainchtein2022user} is responsible for selecting and returning the ingress node IP. After this DNS resolution process, the user sends the request to the ingress node returned in the DNS response. The selected ingress node primarily checks the ``Host'' header and URL in the incoming request and decides whether to serve the request using locally cached content or retrieve the requested content from the origin. In these requests, the ``Host'' header corresponds to a website hosted on CDN to inform the ingress node which website the user wants to access~\cite{cdn_threat,bypass_cdn}.

From the perspective of website owners, to host a website on a CDN and make the ``DNS routing'' work, the website owner can follow a simple procedure. 
Firstly, the website owner registers for CDN services and becomes its customer. Secondly, the website owner should configure the website domain and origin servers on the CDN's customer interface. 
The website domain represents the domain accessed by users, while the origin server indicates where the CDN retrieves the requested resources. Thirdly, the website owner adds DNS records to link the website domain to the CDN.

We can see that domain name play a vital role in CDN service, as a critical option that can be directly configured by the 3rd-party CDN customer, i.e. website owner.
It is well known that, domain name is used to represent a person or organization as an entity on the Internet, an Internet service normally requires its 3rd-party customer to do domain verification procedure to prove the control or ownership over the specific domain name.
However, we have found that many CDN providers lack domain verification or have flawed implementations for domain verification. Consequently, these CDN domain verification problems can lead to security threats, while which has not been systematically analyzed.  

\subsection{Domain Verification}
\label{sec:background-verification}

\noindent \textbf{Domain verification.}
It refers to a series of domain name verification for ensuring the integrity of network traffic and authenticating domain ownership~\cite{sahib-domain-verification-techniques-03}.
It is widely implemented during SSL/TLS certificate issuance, email authentication in transit (using protocols such as SPF~\cite{breakspf}, DKIM~\cite{DKIM}, and DMARC~\cite{DMARC}), and when delegating domain names to third-party hosting providers (e.g., CDNs, DNS, and web hosting services).
Robust domain verification mechanisms play a crucial role in preventing domain names from being abused for malicious purposes.

\begin{figure*}[ht]
  \centering
\includegraphics[width=0.875\linewidth]{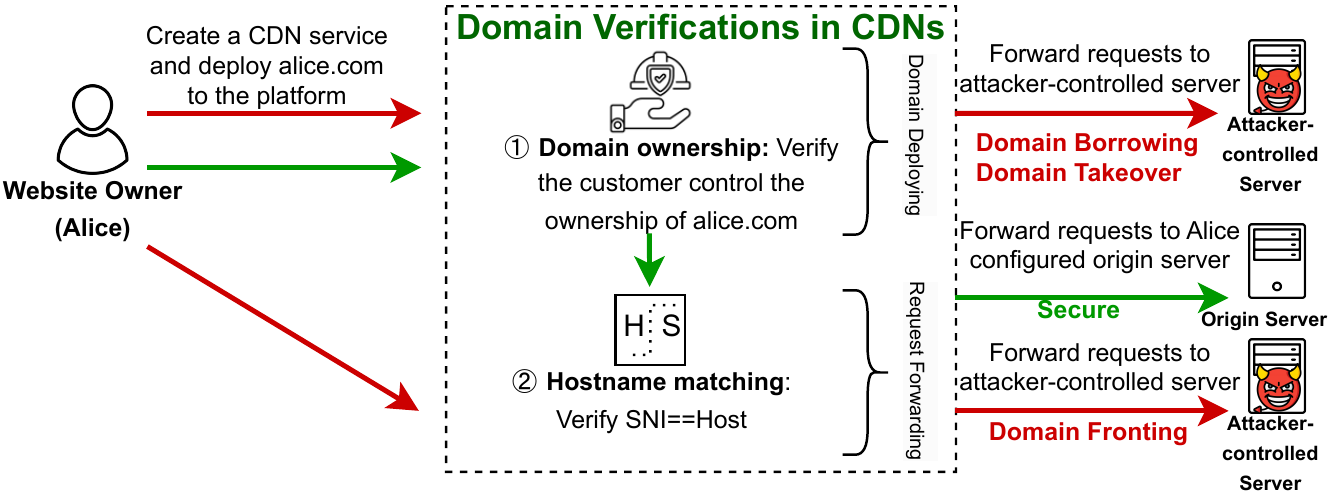} 
  \caption{Domain verification in CDNs.} 
  \label{fig:domain-verification} 
\end{figure*}

\noindent \textbf{Domain verification in CDNs.}
Domain verification in CDN platforms encompasses two critical stages: \textit{\domaindeploying} and \textit{\requestforwarding}, as shown in Figure~\ref{fig:domain-verification}. 
During \textit{\domaindeploying}, CDN platforms should verify that customers have control over configured domain names (steps \ding{192}), ensuring domain ownership verification. This step prevents adversaries from deploying unauthorized domain names and manipulating associated resources.
\lzy{Then, during \textit{\requestforwarding}, CDN platforms need to check hostname matching (step \ding{193}). When HTTPS requests reach CDN ingress nodes, the platform verifies that the Host header matches the Server Name Indication (SNI) in the TLS connection. This verification ensures the CDN forwards the request to the correct origin server.}

\noindent \textbf{Domain verification absence (DVA) vulnerabilities.}
Currently, there are no RFC specifications or widely accepted best practices requiring how CDNs should conduct domain verification.
This lack of standardization can lead to potentially flawed domain verification mechanisms within the CDN ecosystem.
Some research has identified the \textit{Domain Verification Absence} (\textit{DVA}) as a major root cause of domain abuse~\cite{domain_fronting,domain_borrowing_bh,understand_dns_threat}. 
DVA refers to the lack of robust domain verification strategies by CDN platforms, which allows attackers to abuse benign domains hosted on CDNs.
\textit{A CDN is considered to have DVA vulnerabilities if it fails to verify domain ownership during \domaindeploying or neglects hostname matching during \requestforwarding.}

\noindent \textbf{DVA exploitation.}
DVA exploitation manifests in covert communication or domain hijacking. 
Previous research has identified three attack techniques for these scenarios, as outlined in Figure~\ref{fig:domain-verification}.
\lzy{For covert communication, attackers employ methods such as domain fronting~\cite{domain_fronting}, which exploit CDNs that neglect hostname matching during \requestforwarding, and domain borrowing~\cite{domain_borrowing_bh}, which exploit CDNs that do not perform domain verification during \domaindeploying. 
In the case of domain hijacking within CDNs, attackers exploit dangling DNS records~\cite{understand_dns_threat} for domain takeover if CDNs do not perform domain verification or perform flawed domain verification during \domaindeploying.} In a word, DVA exploit opens the door for a series of subsequent complex attacks.

\subsection{Attack Surfaces for DVA}
Below, we introduce how flaws in domain verification during \textit{\domaindeploying} and \textit{\requestforwarding} contribute to the aforementioned three attack surfaces.

\begin{figure}[h]
  \centering
  \includegraphics[width=1.0\linewidth]{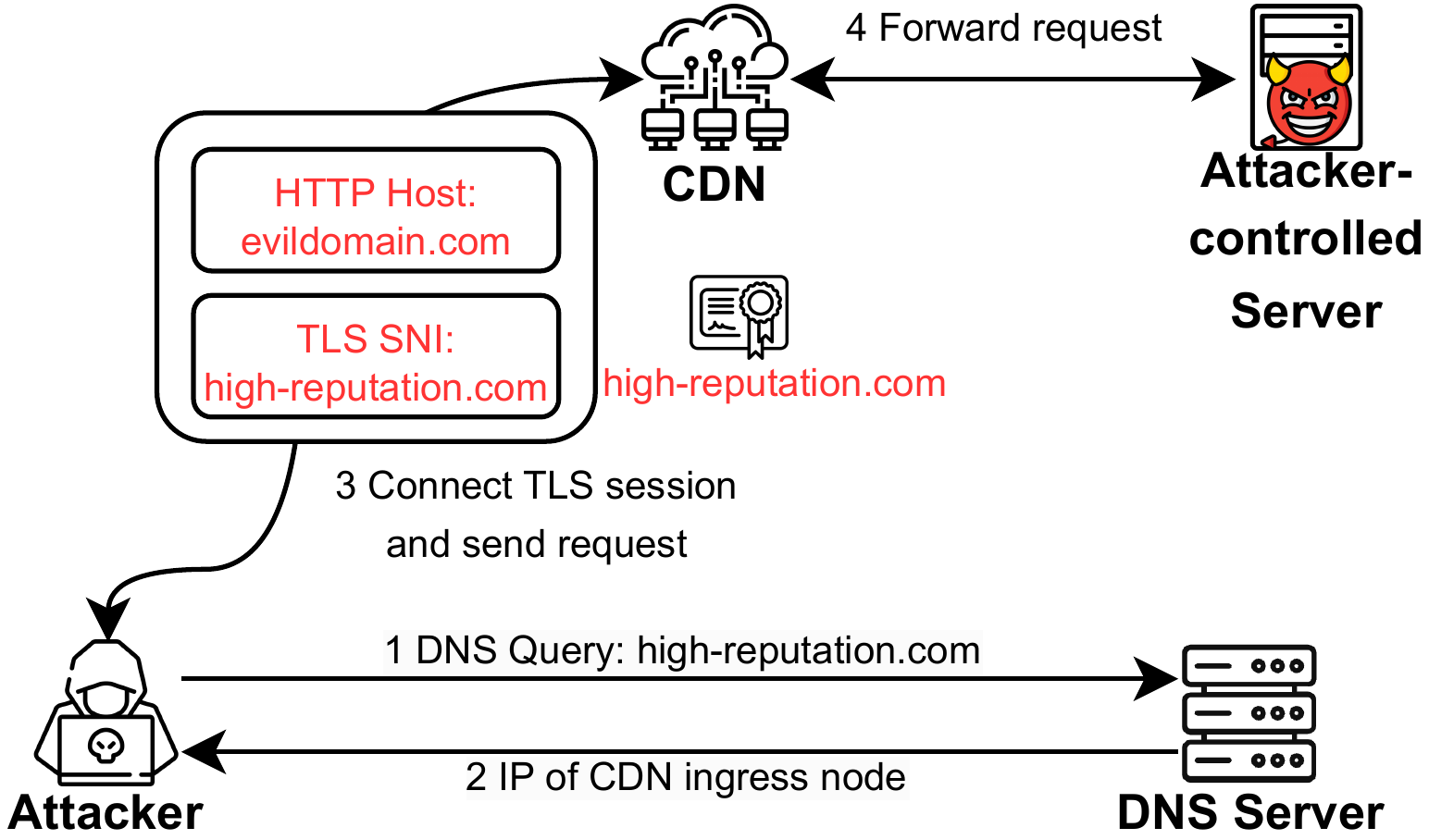} 
  \caption{Domain fronting in the CDN.} 
  \label{fig: dfronting} 
\end{figure}

\noindent \textbf{Domain fronting.}
\lzy{This attack arises when a CDN fails to conduct domain verification during \requestforwarding, allowing it to forward HTTPS requests where the SNI and Host header do not match.}
Assume there's a victim domain named \texttt{high-reputation.com} hosted on a CDN platform. Figure~\ref{fig: dfronting} illustrates how a domain fronting attack operates:
First, an attacker delegates a domain, \texttt{evildomain.com}, on the CDN and configures its origin to point to a attacker-controlled server.
Then, the attacker crafts an HTTPS request to a CDN ingress node. The request sets the SNI as \texttt{high-reputation.com} and the Host header as \texttt{evildomain.com}.
The CDN ingress node, in this scenario, establishes a TLS session using the SNI. However, it forwards the request to the attacker-controlled server configured for \texttt{evildomain.com} based on the Host header. 
This mismatch in hostnames causes what appears to be ``benign'' traffic to reach an attacker-controlled server, potentially facilitating the transfer of malicious traffic.

\begin{figure}[h]
  \centering
  \includegraphics[width=1.0\linewidth]{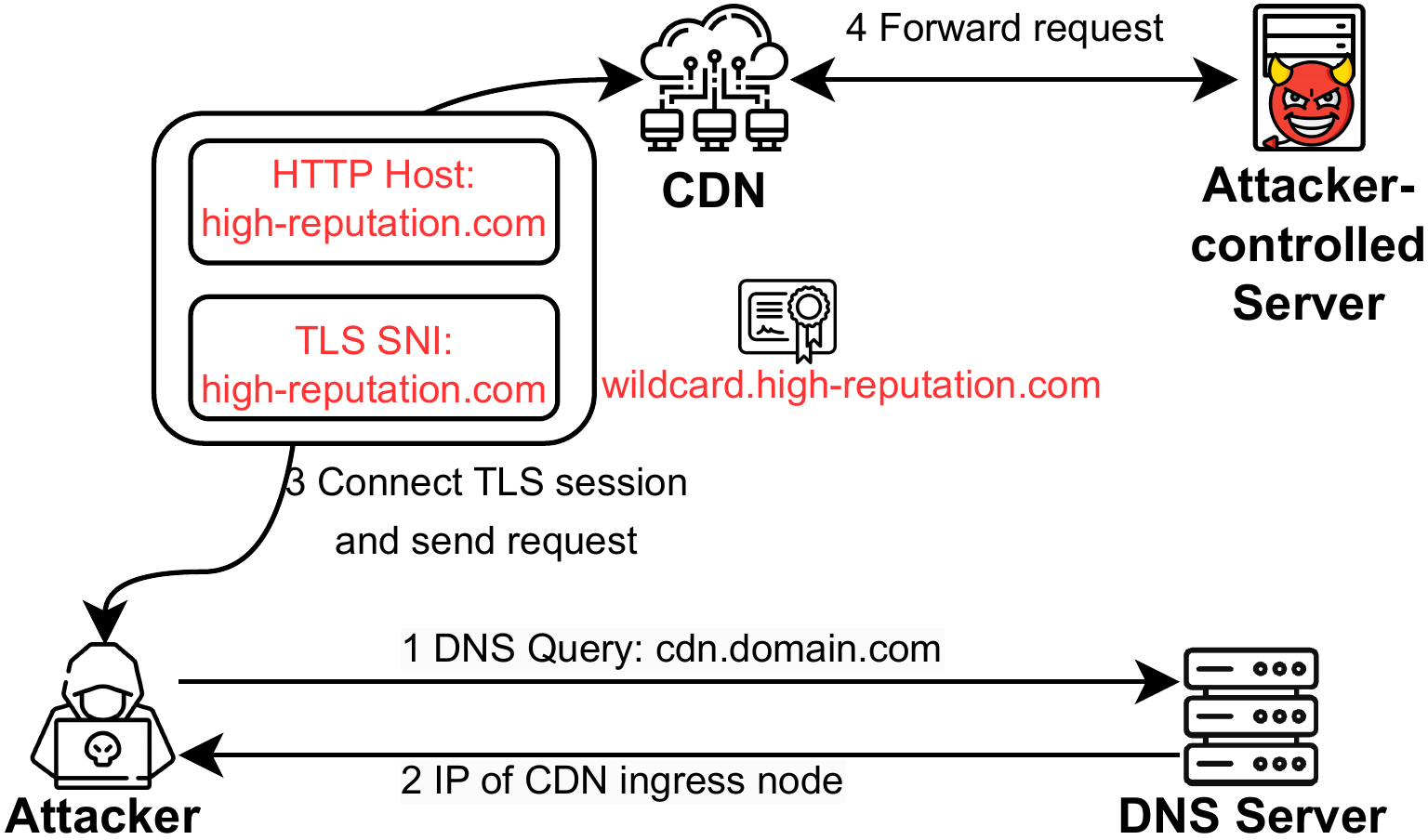} 
  \caption{Domain borrowing in the CDN.} 
  \label{fig: dborrwing} 
\end{figure}

\noindent \textbf{Domain borrowing.}
\lzy{This attack arises when a CDN does not perform domain verification during \domaindeploying, allowing attackers to deploy a high-reputation domain without authority in the CDN.}
Attackers can set the SNI and Host header as high-reputation domain names and directly connect the CDN ingress node. Unlike domain fronting, domain borrowing ensures the consistency of SNI and Host header, which helps to avoid the risk of SSL stripping, thus preventing the risk of being blocked by firewalls. As shown in Figure~\ref{fig: dborrwing}, assuming that an attacker registers \texttt{high-reputation.com} in the CDN, the workflow of domain borrowing is as follows: Firstly, the client performs DNS resolution using another CDN domain name, \texttt{cdn.domain.com}, obtaining the IP address of CDN ingress node. 
\lzy{Secondly, the client sends an HTTPS request setting \texttt{high-reputation.com} as the SNI and Host header to the CDN ingress node.} 
Thirdly, the CDN ingress node searches for the TLS certificate based on the SNI. If it cannot find one, it returns a default CDN shared certificate, such as default.ssl.cdn.com. The client can ignore client certificate validation to establish the TLS connection. Fourthly, the CDN ingress node forwards the HTTPS requests to the attacker-controlled server based on the Host header. Therefore, attackers can hide their traffic in the \texttt{high-reputation.com} domain name.

\begin{figure}[h]
  \centering
  \includegraphics[width=1.0\linewidth]{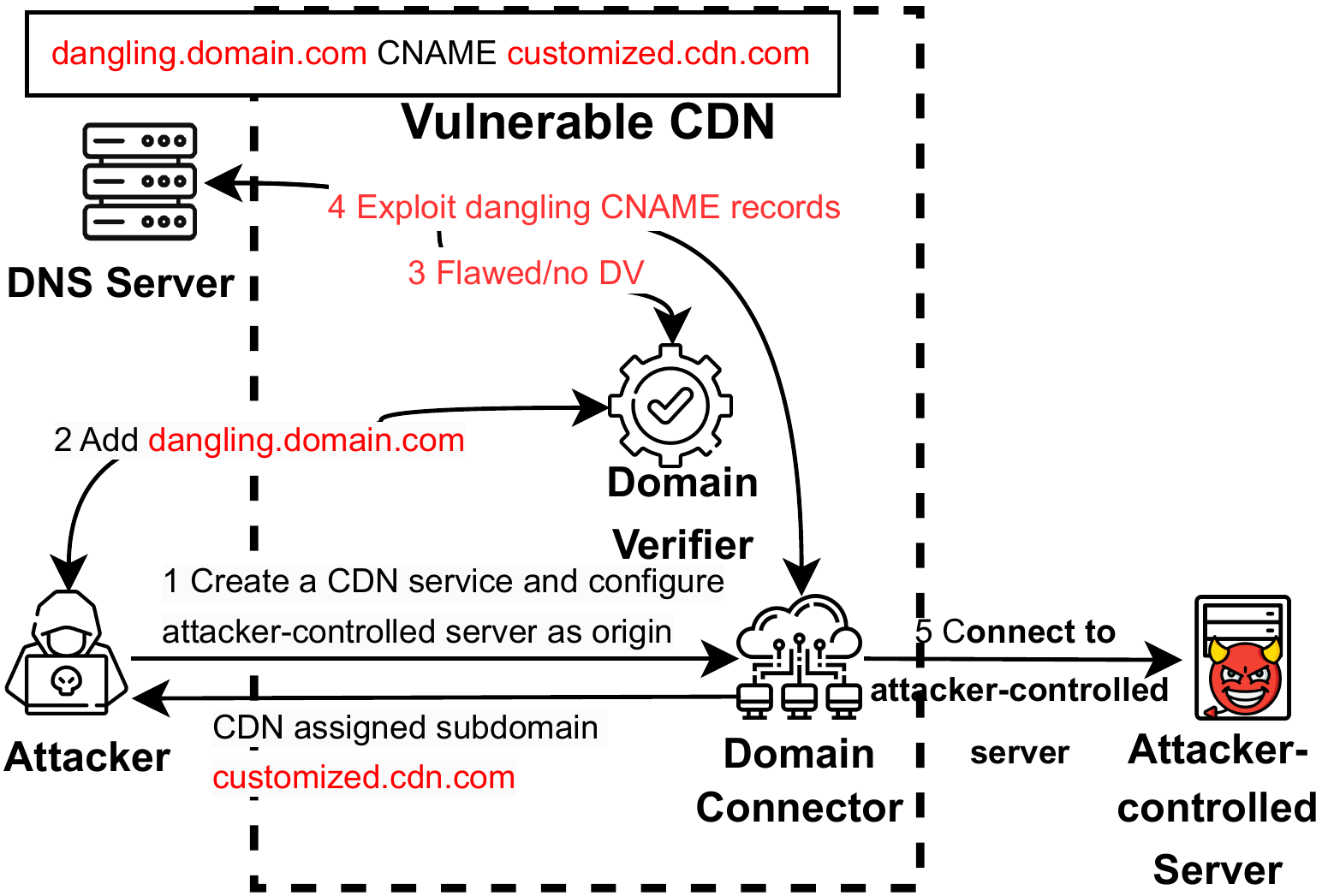} 
  \caption{Domain takeover in the CDN.} 
  \label{fig: dtakeover}
\end{figure}

\noindent \textbf{Domain takeover.}
This attack arises when a CDN does not perform domain verification during \domaindeploying, and the victim forgets to delete DNS records, which allows attackers to exploit the CDN to take over the victim's domain. 
\lzy{When customers terminate CDN services and forget to delete the associated DNS records, it leads to numerous dangling records.
These dangling records still point to CDN-assigned subdomain, which attackers can exploit by registering the same CDN-assigned subdomain. 
As shown in Figure~\ref{fig: dtakeover}, the workflow of domain takeover is as follows\footnote{CDN offers NS and CNAME as two domain deployment methods. CNAME is the most commonly used domain deployment method. The figure illustrates the domain takeover procedure exploiting dangling CNAME records. The domain takeover procedure exploiting dangling NS records is similar.}: If the victim unsubscribes from a CDN service without deleting the associated DNS record. This domain, \texttt{dangling.domain.com}, becomes dangling. Assuming the CDN does not perform domain verification or performs vulnerable domain verification. First, attackers create a CDN service and configure \texttt{attacker-controlled server} as the origin. Second, the attackers add \texttt{dangling.domain.com} to the Domain Verifier. After passing domain verification, the CDN begins offering services to the attacker. Third, the attackers take over \texttt{dangling.domain.com} and carry out further attacks, such as hijacking the website's content or stealing sensitive data.}

\subsection{Comparison with Related Work}
\begin{table*}[t]
    \caption{Comparison of related work}
    \centering
    \small
    \begin{threeparttable}
    \begin{tabular}{lcccccc}
    \toprule[1.5pt]
    Paper          & Scope            & Method & \# Vulnerable CDNs (\%\textsuperscript{1})  & \# Target CDNs & Vulnerable domains\textsuperscript{2} & Across CDNs\textsuperscript{3} \\
    \midrule
    Karthika et al~\cite{measuring_domain_froting} & fronting  & PDNS   & 22 (73.33\%)               & 30            &   \XSolidBrush                   &  \XSolidBrush             \\
    Ding et al.~\cite{domain_borrowing_bh}    & borrowing & manual & 5 (55.56\%)                & 9             &   \XSolidBrush                   &   \XSolidBrush             \\
    Zhang et al.~\cite{dare_shark}   & takeover  & PDNS   & 7 (28.00\%)                & 25            &  \CheckmarkBold                  &  \XSolidBrush             \\
    \toolname            & all              & ADNS   & 43 (95.56\%)               & 45            & \CheckmarkBold                   &   \CheckmarkBold  \\       
    \bottomrule[1.5pt]
    \end{tabular}
    \begin{tablenotes}
    \item[1] \% is the fraction of the vulnerable value to the all value. Take domain fronting for example, 73.3\% (22/30) CDNs are vulnerable.
    \item[2] Vulnerable domains represent whether the system can detect vulnerable domains.
    \item[3] Across CDNs indicate whether the system can detect exploiting a low-security Multi-CDN to abuse a domain hosted by a high-security CDN.
    \end{tablenotes}
    \end{threeparttable}
    \label{table: representative work}
\end{table*}
\lzy{The above vulnerabilities reveal how attackers can exploit CDNs that lack domain verification for covert communication or domain hijacking, leading to subsequent attacks, such as distributing malware or hijacking website content. Therefore, an Internet-scale threat assessment is necessary to understand these potential threats better. However, there is no automated tool that can support Internet-scale assessment.}

The Absence of Domain Verification (DVA) vulnerability is the topic of this study, and the most relevant works are provided in Table~\ref{table: representative work}.
Karthika et al.~\cite{measuring_domain_froting} proposed to measure domain fronting using Passive DNS (PDNS). They measure 30 CDNs and find that 22 CDNs are prone to domain fronting. Compared to them, firstly, \toolname uses Active DNS (ADNS), reducing the usage threshold for security experts without the requirement of obtaining PDNS, which may either be outdated or raise privacy problems. Secondly, as shown in Table~\ref{table: vulnerable CDN overview}, we measure 45 CDNs, find 40 CDNs are prone to domain fronting, and discover 26 new vulnerable CDNs. Thirdly, we extend our target CDN to Multi-CDN and reveal attackers can exploit infrastructure sharing among CDN providers to revive the domain fronting in secure CDN (i.e., CloudFront).
Ding et al.~\cite{domain_borrowing_bh} propose that attackers can leverage domain borrowing for covert C2 communication. They identify 5 out of 9 CDNs as vulnerable to domain borrowing. Compared to them, \toolname automatically discovers more vulnerable CDNs and detects borrowing domains. Secondly, as shown in Table~\ref{table: vulnerable CDN overview}, we measure 45 CDNs, find 24 CDNs vulnerable to domain borrowing, and discover 17 new vulnerable CDNs. Thirdly, we detect 20,296 borrowing domains and uncover two new CDN providers allowing attackers to borrow wildcard certificates.
Zhang et al.~\cite{dare_shark} propose to measure domain takeover using PDNS. They measure 25 CDNs and identify 7 CDNs vulnerable to domain takeover. Compared to them, firstly, \toolname, as shown in Table~\ref{table: vulnerable CDN overview}, we measure 45 CDNs, find 19 CDNs vulnerable to domain takeover, and discover 16 new vulnerable CDNs. Secondly, we analyze the CDN domain verification mechanism and find two new ways of domain takeover. Thirdly, we extend our target CDN to Multi-CDN (i.e., KuaikuaiCloud) and reveal attackers can exploit shared CNAME to takeover dangling domains hosted by secure CDN (i.e., Baidu).

Above all, different from previous research focusing on a single aspect and lacking automatic large-scale evaluation, our research varies from previous works in three key aspects.
Firstly, we systematically explore the CDN domain verification mechanism, discovering two new ways of domain takeover. Secondly, we find that sharing infrastructure across CDNs worsens DVA vulnerability. Finally, previous work measures only a few well-known vulnerable CDNs and has not open-sourced their tools or their tools require PDNS. Therefore, we introduce an automatic detection system, \toolname, using ADNS for large-scale measurement of DVA vulnerability.
\section{Design and Implementation}

\subsection{Motivation}
DVA vulnerability is prevalent in CDNs, posing significant threats to the integrity and security of the Internet. 
Despite efforts by previous researchers ~\cite{domain_fronting,domain_borrowing_bh,takeover_xyz} to mitigate the DVA vulnerability, security incidents continue to occur. APT29 (also known as Cozy Bear) reportedly exploits domain fronting to disguise its C2 (Command \& Control) of malware or botnet infrastructure, thereby complicating detection and attribution \cite{tor_meek}. Another study~\cite{CCTV}  reveals that attackers utilize domain borrowing to bypass network detection, borrowing the reputation of a legitimate domain to distribute RT malware. What is worse, researchers~\cite{Microsoft_takover} have discovered hundreds of dangling Microsoft subdomains, allowing attackers to take them over through Microsoft Azure.

With a comprehensive analysis of previous works, we identify two research gaps in detecting DVA vulnerability. Firstly, there is no universal method to identify vulnerable CDN providers. The CDN providers offered domain connection policies are highly diverse~\cite{hosting_service}, and there are no uniform features to identify vulnerable CDN providers. Secondly, given the large number of CDN providers and the millions of domains, the accurate detection of vulnerable CDN providers and the timely exposure of vulnerable domains poses a significant challenge. In a word, there is still no automatic detection system to identify DVA vulnerability.

\begin{figure*}[t]
    \centering
    \includegraphics[width=1.0\linewidth]
    {images/./figures/crop_system_v17.pdf}
    \caption{\toolname workflow.}
    \label{fig: workflow}
\end{figure*}

\subsection{System Overview}
To fill the research gap in detecting DVA vulnerability, this study designs and implements \toolname to evaluate its severity automatically and periodically at the Internet scale.
As shown in Figure~\ref{fig: workflow}, \toolname consists of three key components: (1) Subdomain Crawler, (2) CDN Checker, and (3) Domain Verification Checker. First, the Subdomain Crawler enumerates subdomains of the targeted domains. 
\lzy{Second, the CDN Checker crawls DNS records, collects ingress node IPs, and discovers domains hosted by CDNs for further domain verification checking. Third, the Domain Verification Checker automatically detects DVA vulnerability based on the information collected by the first two components. }
We will elaborate on the role of each system component.

\subsection{Subdomain Crawler}
We use a custom prefix dictionary to generate a list of Fully Qualified Domain Names (FQDNs) under the Second-Level Domain (SLD), offering greater flexibility in our subdomain discovery process~\cite{subdomain_crawler}. Each prefix is joined with the target SLD to create an FQDN. Then, we validate the existence of these FQDNs using DNS queries. During the verification process, we send query requests to DNS servers to check whether there are corresponding resource records for the FQDNs we created. If the DNS response contains valid records, it indicates that the FQDN exists under the target SLD. To filter out wildcard DNS records, we also include a random prefix in our queries to identify and exclude. Through this approach, we can detect valid FQDNs under the target SLDs and expand the scope of our search.

\subsection{CDN Checker}

We have developed the CDN Checker to provide comprehensive domain data for Domain Verification Checker. The CDN Checker comprises three components that collaborate to conduct thorough domain scanning. These components are (1) Crawling DNS records, (2) Discovering domains hosted by CDNs, and (3) Collecting ingress node IPs.

\noindent \textbf{Crawling DNS records.}
We crawl CNAME, NS, and A records. To accelerate the DNS records crawling for Tranco Top 1M domains, we concurrently launched five instances of our DNS crawler that would start from different indexes in Tranco Top 1M domains. It took approximately one day to finish the scan. 

\noindent \textbf{Discovering domains hosted by CDNs.}
When domain owners want to deploy their domain on a CDN, the CDN assigns a subdomain belonging to the CDN itself. The domain owner must create a CNAME record pointing their domain to the CDN-assigned subdomain to redirect traffic to the CDN. We can identify domains hosted by a CDN by following the steps.
\lzy{Firstly, for a given CDN (e.g., Fastly), we gather CDN-assigned subdomains from a publicly available paper~\cite{cdn_dark_side} and further supplement it using the Public Suffix List~\cite{public_suffix_list}. Secondly, we compare the previously collected DNS records with the CDN-assigned subdomains (see Table~\ref{table: cdn_assigined_subdomain}) to identify the domains hosted on the CDN. Thirdly, we send HTTP requests to the identified domains hosted by the CDN. We then compare their responses with non-CDN-deployed fingerprints (see Table~\ref{table: non-CDN-deployed fingerprints}) to recheck the domains hosted by the CDN. This process ensures the accuracy of our results, clearly distinguishing between CDN-deployed and non-CDN-deployed domains thereby reducing false positives.}

\noindent \textbf{Collecting ingress node IPs.} 
We need CDN ingress node IP addresses to detect domain borrowing. 
The accuracy of these ingress node IP addresses determines the success rate of detecting domain borrowing. Unfortunately, to our knowledge, CDN ingress node IP addresses are not fully publicized on the Internet. Therefore, we need to collect the ingress node IP addresses of CDN by ourselves. Internet-wide scan is a traditional method of collecting the ingress node IPs. Send an HTTP request to the Internet-wide IPv4 space that sets the well-known CDN customers' domain as the Host header. If the response is the correct web page, it indicates the IP is the ingress node IP. However, this method cannot filter out open proxies. We have found that ingress node IPs can be gathered by querying DNS resolvers for CDN customer domains~\cite{measuring_cdn,imc_2013,sigcomm_2015}.
Firstly, we compare the previously collected DNS records of the domain with the CDN-assigned subdomain to gather the domain hosted by CDNs. Secondly, we gather the CDN-hosted domain's A record as the CDN ingress node IPs. Thirdly,  we need to filter these IPs to ensure their validity. We conduct a liveness check by sending HTTP requests and waiting for their responses. If a node fails to respond, responds slowly, or responds without the specific HTTP header (e.g., ``Server: cloudflare''), we consider the IP unavailable and remove it from the list. Fourthly, due to the large size of the CDN, the number of ingress node IPs can be very significant. We classify the ingress node IPs using GeoIP data. Based on this data, we categorize the ingress node IPs according to their city locations. Finally, we randomly select one IP from each city and compile these IPs into a list. This list represents the globally distributed CDN ingress node IPs. 

\begin{table*}[ht]
    \caption{CDN-assigned subdomain.}
    \footnotesize
    \centering
    \begin{threeparttable}
    \begin{tabular}{@{}ll|ll@{}}
    
    \toprule[1.5pt]
    \textbf{CDN}            & \textbf{Assigned subdomain}                                                                     & \textbf{CDN}            & \textbf{Assigned subdomain}                                     \\ 
    \midrule

    \textbf{AligeCDN}          & agilewingcdn.com                                              & \textbf{Goooood}        & .prod.defense-dns.net                             \\
\textbf{Akamai}         & .akamai.net, .akamaiedge.net , .edgesuite.net                               & \textbf{HuaweiCloud}         & .cdnhwc1.com, .cdnhwcxcy07.com.                   \\
 \textbf{Alibaba}        & .cdngslb.com, .alicdn.com, .alikunlun.com, .kunlunsl.com                                        & \textbf{JDCloud}        & .cloud-scdn.com, .jcloud-cdn.com,                 \\
\textbf{ArvanCloud}     & .arvancdn.com, .arvancdn.ir                                                            & \textbf{KeyCDN}         & .kxcdn.com, .kvcdn.com                            \\
\textbf{Azion}          & .map.azionedge.net, .azioncdn.net                                                      & \textbf{KingsoftCloud}       & .ksyuncdn.com, .ks-cdn.com                        \\
\textbf{Azure}          & .azureedge.net, .azurefd.net                                                           & \textbf{KuoCai}         & .kuocaidns.com.                                   \\
\textbf{Baidu}          & .bdydns.com, .jomodns.com, .yunjiasu-cdn.net                                           & \textbf{LeaseWeb}       & .lswcdn.net                                       \\
\textbf{BelugaCDN}      & .belugacdn.com                                                                         & \textbf{Layun}          & .yuncdn.layuncdn.com                              \\
\textbf{Bunny}          & .b-cdn.net, .b-cdn.com, .bunny.net, .bunny.net.                                        & \textbf{LightCDN}       & .r.cdn36.com                                      \\
\textbf{Cachefly}       & .cachefly.net                                                                          & \textbf{Limelight}      & .llnwd.net, .lldns.net                            \\
\textbf{CDN77}          & .rsc.cdn77.org, .cdn77.net, .cdn77-ssl.net                                             & \textbf{Lumen}          & .footprint.net, .fpbns.net                        \\
\textbf{CDNetworks}     & .cdngc.net, .gccdn.net, .qtlgslb.com, .cdnetworks.net                                  & \textbf{Medianova}      & .mncdn.com, .mncdn.org                            \\
\textbf{CDNsun}         & .cdnsun.net                                                                            & \textbf{Netlify}        & .netlify.com, .netlifyglobalcdn.com, .netlify.app. \\
\textbf{CDNvideo}       & .cdnvideo.ru                                                                           & \textbf{Qiniu}          & .qiniudns.com                                     \\
\textbf{ChinaNetCenter} & .qtlcdn.com                                                                            & \textbf{StackPath}      & .hwcdn.net, .stackpathcdn.com, .stackpathdns.com  \\
\textbf{Cloudflare}     & .cdn.cloudflare.net                                                                    & \textbf{Tencent}        & .cdn.dnsv1.com, .tdnsv6.com.                      \\
\textbf{CloudFront}     & .cloudfront.net                                                                        & \textbf{UCloud}         & .ucloud.com.cn                                    \\
\textbf{DogeCloud}      & .s2-web.dogedns.com                                                                    & \textbf{Udomain}        & .xcdn.global                                      \\
\textbf{EdgeNext}       & .bsclink.cn                                                                            & \textbf{UPYun}          & .aicdn.com                                        \\
\textbf{Edgio}          & .glb.edgio.net, .systemcdn.net, .edgecastcdn.net.                                      & \textbf{Yunaq}          & .cdn.jiashule.com                                 \\
\textbf{Fastly}         & .fastly.net, .fastlylb.net                                                             & \textbf{Yundun}         & .cname.hcnamedns.com                              \\
\textbf{G-core}         & .d.gcdn.co                                                                             & \textbf{Sudun}          & .sudun1.suduncdn.com                               \\
    \bottomrule[1.5pt]
    \end{tabular}
    \begin{tablenotes}
    \item[1] KuaikuaiCloud is a Multi-CDN that uses the infrastructure of Tencent or Baidu. So it doesn't have its own subdomain
    \end{tablenotes}
    \end{threeparttable}    
    \label{table: cdn_assigined_subdomain}
\end{table*}

\subsection{Domain Fronting Tester}
Figure~\ref{fig: workflow} illustrates the workflow of the Domain Fronting Tester, which can be divided into three steps: (1) We utilize the previously collected domains hosted by CDNs to identify valid URLs from these domains. (2) We select domains from the same CDN provider and pair them with valid URLs to generate testing tuples. (3) We use these testing tuples to conduct domain fronting testing and verify whether the CDN provider is vulnerable to domain fronting.

\noindent \textbf{Identifying valid URLs served by CDNs.}
After identifying domains hosted by CDN, we crawl the URLs that are actual web objects on these domains. However, a challenge arises when making a direct ``GET /'' HTTPS request to ``https://www.example.com'' without specifying the complete resource path, as it may fail if the web server restricts access to the root directory. Therefore, we have developed a custom Chromium-based web crawler using Selenium~\cite{Selenium}. Our crawler accesses a domain and captures all network requests and corresponding responses. Domain fronting testing requires the content of a URL to remain stable across multiple requests, so we only retain URLs for static web resources such as images, .js, and .css files on www.example.com.

\noindent \textbf{Generating testing tuples.}
For testing domain fronting on a CDN, the process involves selecting a front domain (called \textit{Fd}), a target domain (called \textit{Td}) also hosted by the same CDN provider, and a URL from the target domain (called \textit{Ut}). The steps are: (1) establish a TLS connection using the \textit{Fd} as the Server Name Indication (SNI), (2) send an HTTPS request for the \textit{Ut} with the Host header set to the \textit{Td}. If the web object is fetched successfully without errors and the SNI remains set to \textit{Fd}, the CDN is vulnerable to domain fronting. However, in practice, the above process is not sufficient. It's also necessary to verify that the object returned from the HTTPS request is identical to the original web object without SNI alteration. Furthermore, we need to ensure consistent results across different domain pairs (\textit{Fd}, \textit{Td}) within the CDN to determine if the entire infrastructure is vulnerable to domain fronting. Therefore, we select up to 10 tuples (\textit{Fd}, \textit{Td}, \textit{Ut}) from the CDN. For each tuple, the \textit{Fd} serves as the SNI, the \textit{Td} hosts the \textit{Ut}, and both domains are under the same CDN provider. 

\noindent \textbf{Recognizing vulnerable CDNs.}
We processes each selected tuple (\textit{Fd}, \textit{Td}, \textit{Ut}) as follows:

\begin{itemize}

\item[$\bullet$] \textit{Step 1:} Request target URL with target domain as SNI and Host header.
We craft an HTTPS request for the URL \textit{Ut}, setting both the Host header and SNI as \textit{Td}. We store the response content as \textit{Rt}.

\item[$\bullet$] \textit{Step 2:} Request target URL with front domain as SNI and target domain as Host header. We craft an HTTPS request for \textit{Ut} but set the SNI as \textit{Fd} while setting the Host header as \textit{Td}. We store the response content as \textit{Rv}.

\item[$\bullet$] \textit{Step 3:} Request target URL with front domain as SNI and Host header. We craft an HTTPS request for \textit{Ut}, setting both the Host header and SNI as \textit{Fd}. This step verifies that \textit{Ut} is not available under \textit{Fd}, ensuring the test's validity. We record the response content as \textit{Rf}.

\end{itemize}

By analyzing the responses in the following steps, we determined that the domain fronting tests were successful:

\begin{itemize}
\item[$\bullet$] \textit{Rt} is a valid HTTP response without errors.
\item[$\bullet$] \lzy{\textit{Rv} should match \textit{Rt}, indicating consistent behavior of the target URL when the SNI setting is the front domain, but the Host header set is the target domain.}

\item[$\bullet$] \textit{Rf} should be either empty (indicating response 404 Not Found) or should be different from \textit{Rt}, confirming that the front domain does not serve the content of the target URL when both the SNI and the Host header set as the front domain.

\end{itemize}

To compare the content of \textit{Rt}, \textit{Rv}, and \textit{Rf}, we compute and compare their SHA1 hashes~\cite{eastlake2001us}. We repeat these tests up to ten times for each CDN, with each test based on a randomly chosen tuple. After these processes, we are able to recognize all the vulnerable CDN providers.

\subsection{Domain Borrowing Finder}
Figure~\ref{fig: workflow} shows the workflow of Domain Borrowing Finder, which can be divided into three steps: (1) We identify vulnerable CDNs and collect responses of the domain non-CDN-deployed as non-CDN-deployed fingerprints. (2) We use the previously collected domains not hosted by CDN as Host headers to generate legal HTTP requests. (3) We send generated HTTP requests to CDN ingress nodes and compare the HTTP response with non-CDN-deployed fingerprints to identify the borrowing domains.

\noindent \textbf{Identifying vulnerable CDNs.}
To identify CDN providers vulnerable to domain borrowing, we need to collect responses of the non-CDN-deployed domain. Although previous work~\cite{domain_borrowing_bh} has manually identified that five CDN providers are vulnerable to domain borrowing, there have been no systematically explored CDN providers and collected non-CDN-deployed fingerprints. Therefore, we adopt the following automated approach:

\begin{itemize}
\item[$\bullet$] \textit{Step 1:} 
\lzy{Domain verification means the person who owns the domain and can manage the domain's authoritative DNS server.}
Customers can claim domain ownership on a CDN provider in two ways. (1) DNS-based verification: CDN providers generate a challenge token and ask customers to configure it in a DNS record (e.g., CNAME or TXT). (2) Web-based verification: CDN providers ask customers to upload a file containing a challenge token to a specific directory on the website. Many Internet services now need domain verification, but there are no standard practices. We inspect their domain verification mechanism as follows. Firstly, we review CDN providers' operational documentation and search for service setup tutorial videos to find the officially claimed domain verification mechanism. \lzy{Secondly, we will register one test account for each CDN provider, create a service, and configure our server as origin.
We inspect whether the test account can set the custom domain as a high-reputation domain that does not belong to us.}

\item[$\bullet$] \textit{Step 2:} We send an HTTP request to the CDN ingress node IP with the custom domain set in CDN as the Host header. If the response contains the correct webpage, it indicates the CDN is vulnerable to domain borrowing.

\item[$\bullet$] \textit{Step 3:} \lzy{For CDN providers vulnerable to domain borrowing, we send an HTTP request to the ingress node IP with a randomly generated non-CDN-deployed domain as the Host header.} We extract the unique HTTP headers and HTTP body from its HTTP response as the non-CDN-deployed fingerprints (as shown in Table~\ref{table: non-CDN-deployed fingerprints}).

\end{itemize}

\noindent \textbf{Generating HTTP requests and recognizing borrowing domains.}
Firstly, we use the previously collected domains not hosted by CDN as the Host headers to generate HTTP requests. Secondly, we send HTTP requests to the ingress node IPs. However, to avoid burdening the CDN services with the following measuring, we select one ingress node IP from each city as a set representing all ingress node IPs. 
Thirdly, we compare HTTP responses with non-CDN-deployed fingerprints. A mismatch indicates that the domain is engaged in borrowing.

\begin{table}[h]
\caption{Non-CDN-deployed fingerprints.}
\footnotesize
\begin{threeparttable}
\begin{tabular}{lcl}
\toprule[1.5pt]
\textbf{CDN}            & \textbf{Code\textsuperscript{2}} & \textbf{Fingerprint\textsuperscript{3}}           \\ 
\midrule
Azion          & 404              & B: Not Found                              \\
Bunny          & 403              & B: Domain suspended or not configured          \\
Cachefly       & 403              & B: Hostname not configured                     \\
CDN77          & N/A              & N/A                                         \\
CDNetworks     & N/A              & N/A                                         \\
CDNsun     & 400              & B: 400 Bad Request                                         \\
ChinaNetCenter & N/A              & N/A                                         \\
CloudFront     & 403              & B: The request could not be satisfied   \\
DogeCloud      & 404              & H: X-Cache-Lookup: Return Directly             \\
EdgeNext       & 403              & B: ERROR: ACCESS DENIED                        \\
Edgio          & 404              & B: 404 - Not Found                             \\
Fastly         & 500              & B: Fastly error: unknown domain           \\
Goooood        & 400              & B: /unkonwdomain404/notfound                   \\
KeyCDN         & 403                 & B: the resource has been denied \\
KuoCai         & 403                 & H: Byte-Error-Code: 0060 \\
Layun          & 200              & B: Error in website request\textsuperscript{4} \\
LightCDN       & 400              & B: 400 Bad Request                             \\
Medianova      & N/A              & N/A                                         \\
Netlify        & 404              & B: Not Found - Request ID                     \\
StackPath      & N/A              & N/A                                            \\
Sudun          & 200              & B: Please use the domain name to access\textsuperscript{4} \\ 
UCloud         & 403              & B: ERROR: ACCESS DENIED                        \\
Udomain        & 503              & B: UDomain CDN | Error 503                     \\
Yundun         & N/A              & N/A                                         \\
\bottomrule[1.5pt]
\end{tabular}
\begin{tablenotes}
\item [1] N/A means that the CDN does not respond to the request.
\item [2] Code means HTTP response status codes.
\item [3] Fingerprint types: B=HTTP Body, H =HTTP Header.
\item [4] In most cases, the returned fingerprits are in English.
\end{tablenotes}
\end{threeparttable} 
\label{table: non-CDN-deployed fingerprints}
\end{table}

\subsection{Domain Takeover Detector}
Figure~\ref{fig: workflow} shows the workflow of Domain Takeover Detector, which can be divided into three steps: (1) We identify vulnerable CDNs and collect HTTP responses of the domain stopped its CDN service as CDN-service-discontinued fingerprints. (2) We use the previously collected domain data to discover domains hosted by vulnerable CDNs. (3) We send HTTP requests to domains hosted by vulnerable CDNs and compare HTTP responses with CDN-service-discontinued fingerprints.

\noindent \textbf{Identifying vulnerable CDNs.}
To identify CDNs vulnerable to domain takeover, we investigated their domain connection and domain verification mechanism. Firstly, we reviewed the documentation provided by CDN providers and searched for tutorial videos on service configurations to understand the officially stated domain verification mechanism. 
\lzy{Secondly, we registered two test accounts\footnote{We registered two test accounts to verify that the CDN vendors did not perform the domain verification mechanism or that they performed the flawed domain verification mechanism. This only confirms DVA vulnerability and does not impact the automation of \toolname.} for each CDN provider to validate the feasibility of domain takeover in practice. Step 1: we use an account (as a victim) to create a CDN service and add DNS records to deploy a domain in CDN. Step 2: we delete the CDN service to make the victim's domain dangling. Step 3: we use another account (as an attacker) to create a CDN service and attempt to take over the victim's dangling domain.}
Previous research~\cite{understand_dns_threat,takeover_xyz,dare_shark} has primarily focused on widely used web-hosting services, but systematic testing of CDN has not been fully conducted. To fill this gap, we systematically explored the practices currently adopted by CDNs and identified 19 vulnerable CDNs. 
\lzy{Additionally, we discovered new vulnerabilities in the domain verification mechanism. Details of exploiting these vulnerabilities for domain takeover are in Section~\ref{subsec: Newfindings}.}
When a CDN service is deactivated, the CDN responds to the client with distinctive HTTP or DNS responses. We collected these responses as CDN-service-discontinued fingerprints to determine service status and detect dangling domains. Throughout the research process, we summarized three types of fingerprints (in Table~\ref{talbe: CDN service discontinuation fingerprints}):

\begin{itemize}

\item[$\bullet$]HTTP response headers. 
These are essential for assessing the status of a domain. CDNs typically use default HTTP error codes, like ``404 Not Found'' to indicate the unavailability of services. To distinguish between service discontinuation and other types of failures, we also extract specialized HTTP headers from each service alongside status codes. 

\item[$\bullet$]HTTP response bodies. 
CDNs often employ default error pages with similar HTML structures or specific notification phrases to indicate service status. We utilize the typical content of these pages as fingerprints, such as the phrase ``Fastly error: unknown domain'' to identify service discontinuation, following methodologies from previous research~\cite{takeover_xyz}.

\item[$\bullet$]DNS responses.
CDNs may deliver customized DNS responses when a domain's service is canceled. For instance, domains might resolve to specific IP addresses, such as 127.0.0.1. In cases where the service is discontinued and caches have expired, assigned sub-domains may return non-existent domain errors (i.e., NXDOMAIN).

\end{itemize}

\noindent \textbf{Discovering domain hosted by vulnerable CDNs and recognizing dangling domains.}
Firstly, to discover domains hosted by vulnerable CDNs, the Domain Takeover Detector analyzes domains' DNS records to verify whether they match the fingerprints of CDN providers previously identified as vulnerable to domain takeover. 
Secondly, the Domain Takeover Detector compares the DNS responses provided by CDN Checker with CDN-service-discontinued fingerprints. If there is no match, the Domain Takeover Detector sends HTTP requests to the domain and compares the HTTP responses with  CDN-service-discontinued fingerprints. Ultimately, domains with matched CDN-service-discontinued fingerprints are considered dangling domains.

\begin{table}[ht]
\caption{CDN-service-discontinued fingerprints.}
\footnotesize
\begin{threeparttable}
\begin{tabular}{lcl}
\toprule[1.5pt]
\textbf{CDN}                     & \textbf{Code\textsuperscript{2}} & \textbf{Fingerprint\textsuperscript{3}} \\
\midrule
Azure        & NA              & D: NXDOMAIN         \\
Bunny         & 403              & B: Domain suspended or not configured       \\
Cachefly       & 404              & B: hostname not configured                   \\
CDNetworks     & NA               & D: NXDOMAIN                                 \\
ChinaNetCenter & NA               & D: NXDOMAIN                                 \\
Cloudflare     & 530                 & \makecell[l]{ B: Cloudflare is currently unable \\to resolve your requested domain}                                            \\
DogeCloud          & 404              & H: X-Cache-Lookup: Return Directly                             \\
EdgeNext       & NA               & only have one A record.                     \\
Edgio        & 404              & H\&B: 404 - Not Found                             \\
Fastly        & 500              & B: Fastly error: unknown domain          \\
G-core         & NA               & D: SERVFAIL                              \\
KuaikuaiCloud         &NA                & D: 127.0.0.1                                \\
KuoCai         & NA               & D: NXDOMAIN                                 \\
Layun          & 200              & B: Error in website request\textsuperscript{4} \\
LightCDN       & NA               & D: NXDOMAIN                                 \\
Netlify        & 404              & B: Not Found - Request ID                   \\
Sudun          & 200              & B: Please use the domain to access\textsuperscript{4} \\ 
UCloud         & NA               & D: NXDOMAIN                                 \\
Yundun         & NA               & D: NXDOMAIN                                 \\
\bottomrule[1.5pt]
\end{tabular}

\begin{tablenotes}
\item [1] N/A means that the CDN does not respond to the request.
\item [2] Code means HTTP response status codes.
\item [3] Fingerprint types: D=DNS Response, B=HTTP Body, H=HTTP Header.
\item [4] In most cases, the returned fingerprints are in English.
\end{tablenotes}
\end{threeparttable} 
\label{talbe: CDN service discontinuation fingerprints}
\end{table}

\section{MEASUREMENT AND FINDINGS}
\subsection{Domain Analysis Results}
The high-ranking domains hosted on CDNs play a crucial role in DVA vulnerability. When the CDN provides services to high-ranking domains, the ingress node IP addresses may be considered benign and allowed by network security policies, even if shared by multiple domains. For domain fronting, leveraging high-ranking domains as SNI to make TLS connections with CDN ingress nodes can reduce the risk of being blocked. 
It has been proven beneficial for actors, both malicious and benign, who utilize domain fronting as a means to disguise their traffic and evade detection. According to the research by Ding et al.~\cite{CCTV}, attackers can leverage wildcard certificates for domain borrowing when they don't upload the certificates for the domain and ensure that the SNI is the same as the Host header, further reducing the risk of being blocked. 
They demonstrate that leveraging domain borrowing to register a non-existent domain can bypass Palo Alto Firewall detection~\cite{kokko2017next}. 
For domain takeover, attackers can hijack dangling high-ranking domains to carry out malicious activities such as phishing and fraud. Furthermore, Marco et al.~\cite{squarcina2021can} found that attackers can exploit domain takeover to launch Same-Site attacks. 

\begin{figure}[ht]
    \centering
    \includegraphics[width=1.0\linewidth]
    {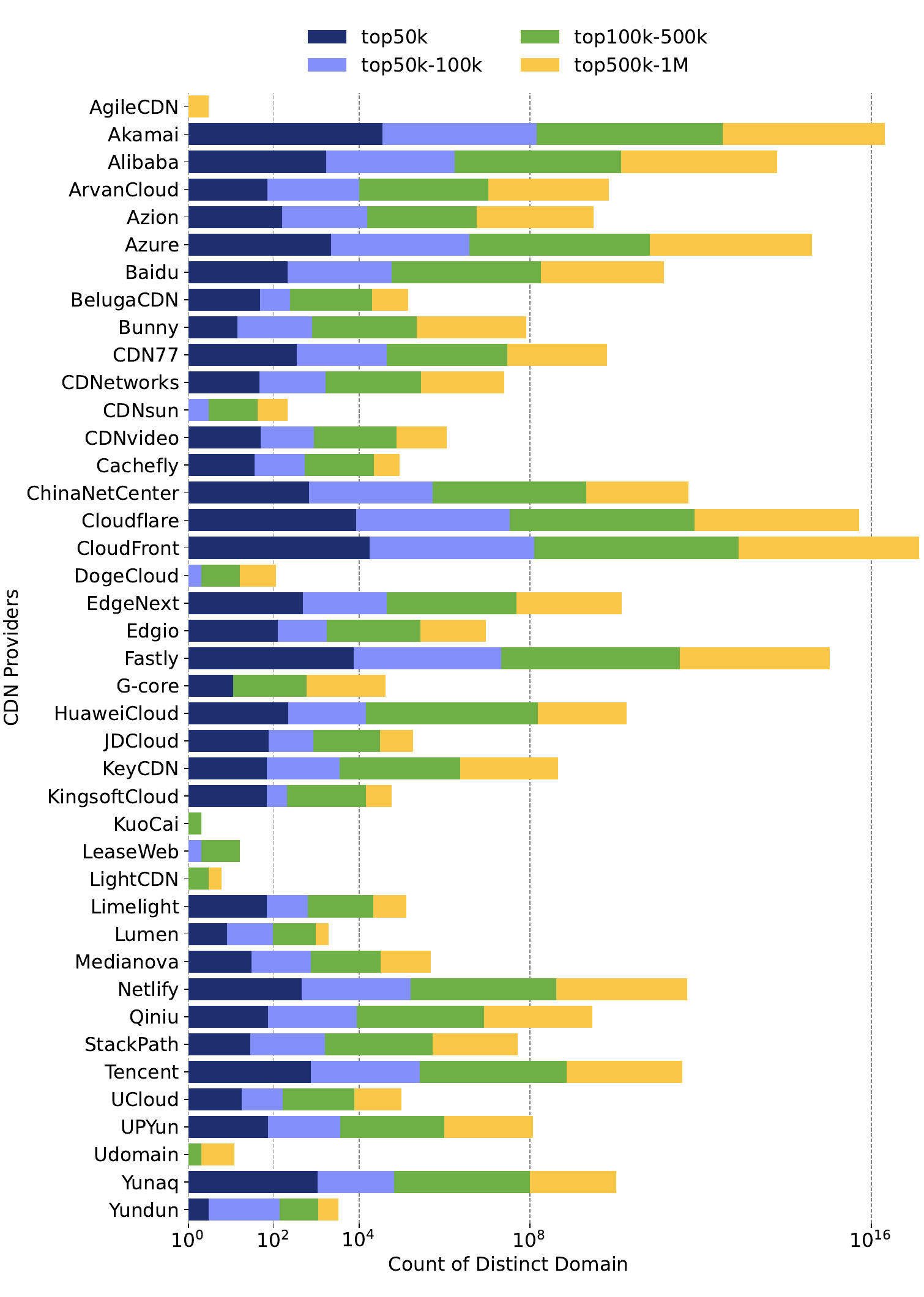}

    \caption{Count of domains per CDN and their Tranco ranking.}
    \label{fig: domain_hosted_by_cdns}
\end{figure}

Therefore, we also explore the distribution of popular (i.e., high-ranking) domains across the different CDN providers. Figure~\ref{fig: domain_hosted_by_cdns} shows the distribution of popular domains in different Tranco ranking ranges covered by each CDN provider. Surprisingly, 37 CDN providers serve popular domains ranked Top 50k. Other CDN providers also cover popular domains ranked Top 500k. The experimental results show that high-ranking domains are not only hosted by well-known CDNs (such as Akamai, Cloudflare, etc.). In contrast to common belief, some high-ranking domains are hosted by lesser-known CDN providers (such as EdgeNext, UCloud, etc.).

\subsection{Vulnerable CDN Providers Overview.}
As illustrated in Table~\ref{table: vulnerable CDN overview}, \toolname has identified 43 CDN providers affected by DVA vulnerability. Our findings reveal that 40 CDN providers are prone to domain fronting, including 26 CDN providers that were previously undiscovered. We also found 24 CDN providers vulnerable to domain borrowing, including 17 previously undiscovered CDN providers, marking a four-fold increase in the number of vulnerable providers compared to previous studies~\cite{domain_borrowing_bh}. 
Numerous studies on domain takeover have primarily focused on a few well-known CDNs, limiting the scope of detection. In contrast, \toolname expanded its scope of detection to 45 CDN providers, including Multi-CDN providers. We find 19 CDN providers vulnerable to domain takeover, including 16 previously unidentified as vulnerable, which is twice the number of vulnerable providers in the previous study.

\begin{table}[ht]
    \caption{DVA vulnerable CDN overview.}
    \small
    \begin{tabular}{lccc}
    \toprule[1.5pt]
   &  \# Vulnerable (\%\*)     & \# Newly Discovered \\
        
    \midrule
    Domain fronting  &   40 (88.88\%)                                  & 26                 \\
    Domain borrowing &  24 (53.33\%)                                   &  17                \\
    Domain takeover  &   19 (42.22\%)                                &  16               \\
    \bottomrule[1.5pt]
    \end{tabular}
    \label{table: vulnerable CDN overview}
\end{table}

\noindent \textbf{Domain fronting results.}
We discovered that 40 CDN providers are vulnerable to domain fronting, even some of the most popular ones like Akamai and Fastly. 
To ensure accurate results from automated tests, we conducted multiple test cases with different parameters for each CDN. We generated tuples for testing domain fronting on each CDN. However, since the number of all possible tuples we could create for each CDN was significantly large, we set limits on the number of domain and URL combinations used for testing. It can prevent any significant load on the CDN infrastructure. Specifically, we randomly selected up to 10 domains per CDN, and for each domain, we randomly chose up to 10 URLs.

\begin{figure}[ht]
    \centering
    \includegraphics[width=1.0\linewidth]
    {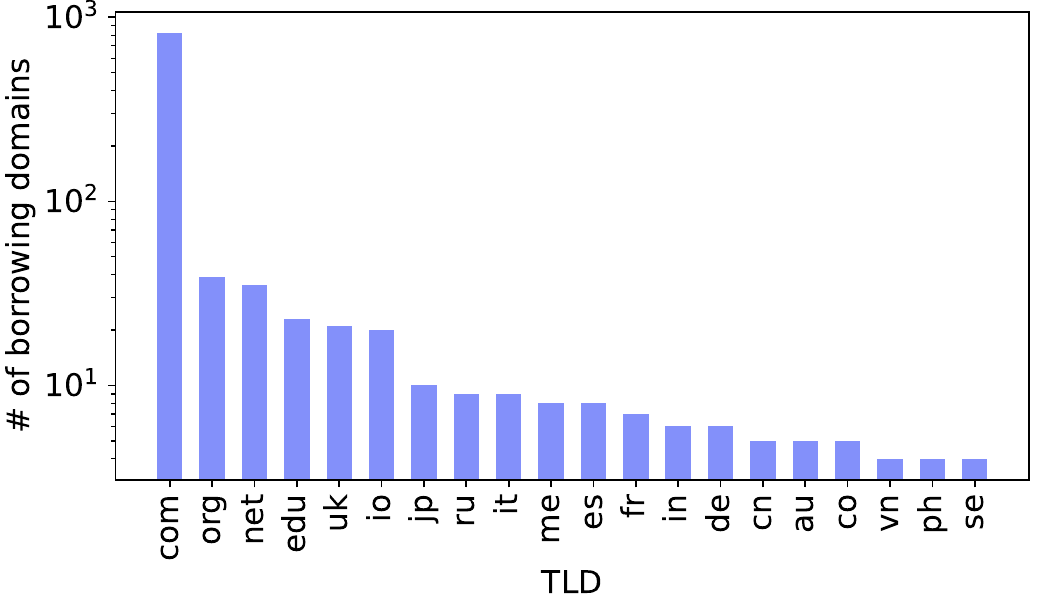}
    \caption{Number of borrowing domains aggregated by TLD.}
    \label{fig: borrowing_TLD}
\end{figure}

\noindent \textbf{Risk and prevalence of borrowing domains.} 
Attackers prefer to borrow high-ranking domains that are easier to bypass firewalls and censorship systems, so we only measure the Tranco Top 10k domains (SLDs) to assess the risk and prevalence of borrowing domains. We discover 20,296 borrowing subdomains under the Tranco Top 10k domains. We aggregate the 20,296 borrowing subdomains by top-level domain (TLD) and present the top 20 TLDs in Figure~\ref{fig: borrowing_TLD}. The top TLD is com, which accounts for 71.8\% of borrowing subdomains. Of particular interest, the TLDs org and edu are considered more trustworthy domains~\cite{org_more_trust,edu_more_trust}, so attackers also like to borrow them. Attackers could exploit the trust in org and edu domains for attacks like bypassing firewalls.

Ding et al.~\cite{domain_borrowing_bh} proposed attackers can borrow the wildcard certificates hosted in CDNs. In order to understand the risks of domain borrowing, we experimented on all vulnerable CDN providers. As shown in Table~\ref{talbe: borrowing cert}, experiment results indicate that 15 out of the 21 CDN providers allow borrowing of the CDN's shared certificates (such as default.fastly.ssl.net) hosted on the CDN. 
Only 4 CDN providers have a vulnerability in the certificate matching mechanism that attackers can exploit to borrow subdomains(e.g., sub.example.com) and wildcard certificates(e.g., *.example.com) from other users.

\begin{table}[ht]
    \caption{Three type of domain borrowing.}
    \centering
    \footnotesize
    \begin{tabular}{@{}lccc@{}}
    \toprule[1.5pt] 
    \multirow{2}[1]{*}{\textbf{CDN Provider}} &
      \multicolumn{2}{c}{\textbf{\# HTTPS}} &
      \textbf{\# HTTP}
      \\
    \cmidrule{2-4}

        & shared certificate & wildcard certificate & None    \\ 
    \midrule

    Azion          & \CheckmarkBold                 &                    & \CheckmarkBold        \\
    Bunny          & \CheckmarkBold                 &                     &  \CheckmarkBold       \\
    Cachefly       & \CheckmarkBold                 & \CheckmarkBold                   &  \CheckmarkBold       \\
    CDN77          & \CheckmarkBold                 & \CheckmarkBold                     & \CheckmarkBold        \\
    CDNetworks     &                    &                      & \CheckmarkBold      \\
    CDNsun         & \CheckmarkBold            &                      & \CheckmarkBold        \\
    ChinaNetCenter &                    &                      & \CheckmarkBold      \\
    CloudFront & \CheckmarkBold                   &                      & \CheckmarkBold       \\
    Dogecloud      & \CheckmarkBold                 &                      & \CheckmarkBold        \\
    EdgeNext       & \CheckmarkBold                 &                     &\CheckmarkBold         \\
    Edgio          & \CheckmarkBold                 &                      &\CheckmarkBold         \\
    Fastly         & \CheckmarkBold                 &                      &\CheckmarkBold         \\
    Goooood        &                    &                &\CheckmarkBold \\
    KeyCDN         & \CheckmarkBold            &                      &\CheckmarkBold         \\
    KuoCai         &                    &                      & \CheckmarkBold      \\
    Layun          &                    &                      & \CheckmarkBold      \\
    LightCDN       & \CheckmarkBold                 &                      & \CheckmarkBold         \\
    Medianova      & \CheckmarkBold                   &                      &\CheckmarkBold  \\
    Netlify        & \CheckmarkBold                 & \CheckmarkBold                   &\CheckmarkBold         \\
    StackPath      & \CheckmarkBold                   & \CheckmarkBold                     &\CheckmarkBold  \\
    Sudun          &                    &                      & \CheckmarkBold      \\
    UCloud         & \CheckmarkBold                 &                      & \CheckmarkBold        \\
    Udomain        &                    &                      & \CheckmarkBold      \\
    Yundun         &                    &                      & \CheckmarkBold      \\ 
    \bottomrule[1.5pt]
    \end{tabular}
    \label{talbe: borrowing cert}
\end{table}

\begin{figure}[ht]
  \centering
  \includegraphics[width=1.0\linewidth]{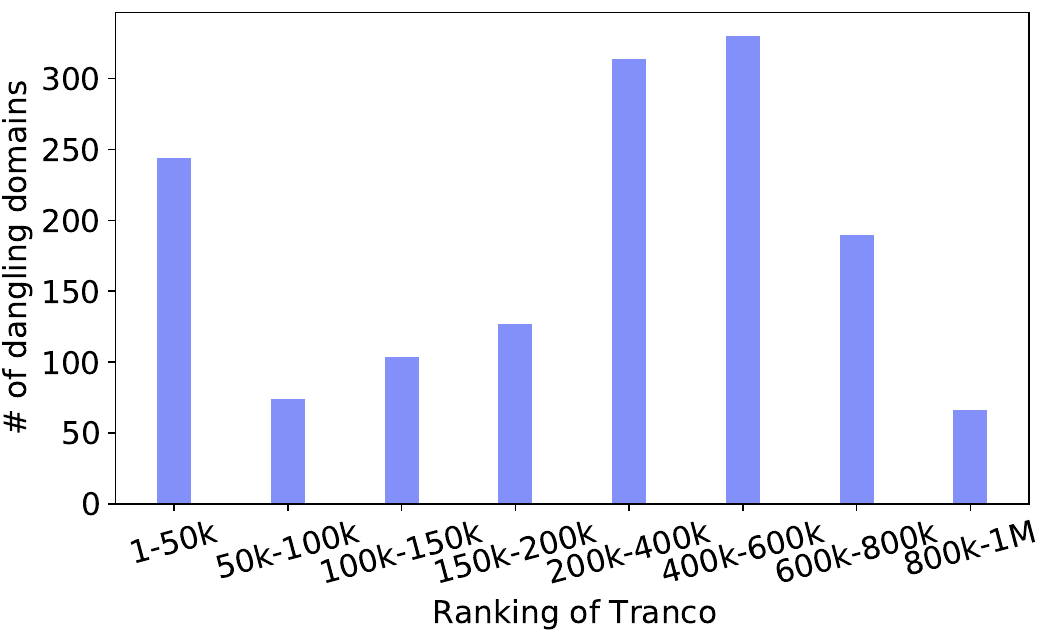} 
  \caption{Tanco rank distribution for the dangling domains.} 
  \label{fig: dangling rank} 
\end{figure}

\noindent \textbf{Risk and prevalence of dangling domains.}
We measure the FQDNs of Tranco Top 1M domains using \toolname to understand the risk and prevalence of dangling domains. The results show that Tranco Top 1M domains have 1449 dangling domains. Figure~\ref{fig: dangling rank} shows the distribution of Tranco rankings for these dangling domains. Of these, 244 domains are in the Tranco Top 50k.

\begin{figure}[ht]
    \centering
    \includegraphics[width=1.0\linewidth]
    {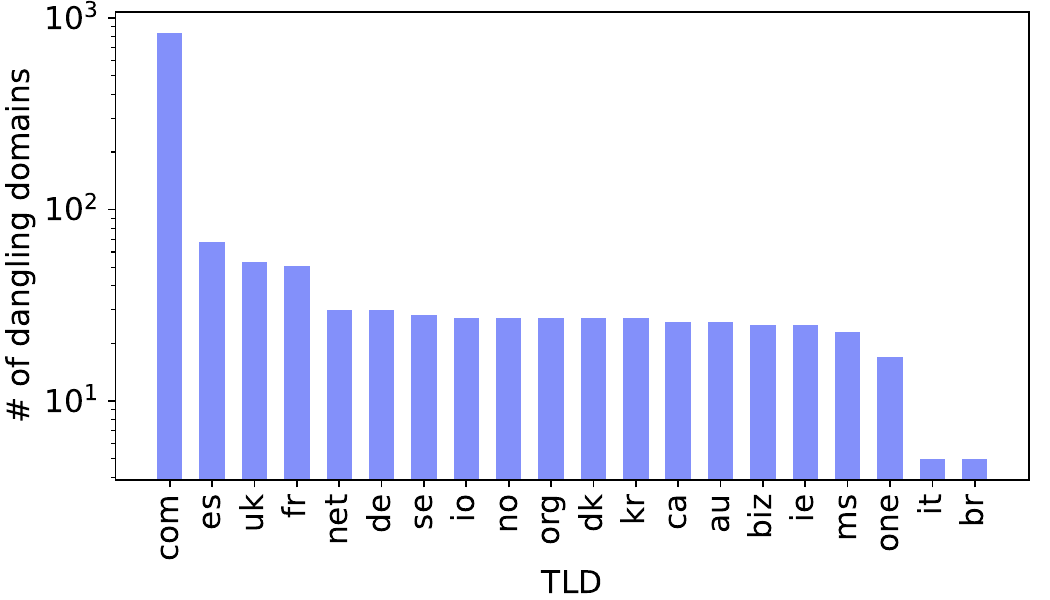}
    \caption{Number of dangling domains aggregated by TLD.}
    \label{fig: dangling_TLD}
\end{figure}

To understand the risks of these dangling domains, we aggregate the 1449 dangling domains by TLD and present the top 20 TLDs in Figure~\ref{fig: dangling_TLD}. The top TLD is com, which accounts for 57.48\% of dangling domains. 
Of particular interest, the TLDs org and edu are well-managed DNS zones, adhering to eligibility requirements and a strict process for registering new domains. However, they still have 27 and 1 dangling domains.
As a result, attackers could exploit the trust in org and edu for attacks like phishing and scams.

\subsection{New findings}\label{subsec: Newfindings}

\begin{figure*}[ht]
    \centering
    \subfigure[Domain takeover by exploiting CDN domain misconnection.]{\includegraphics[width=0.488\textwidth]{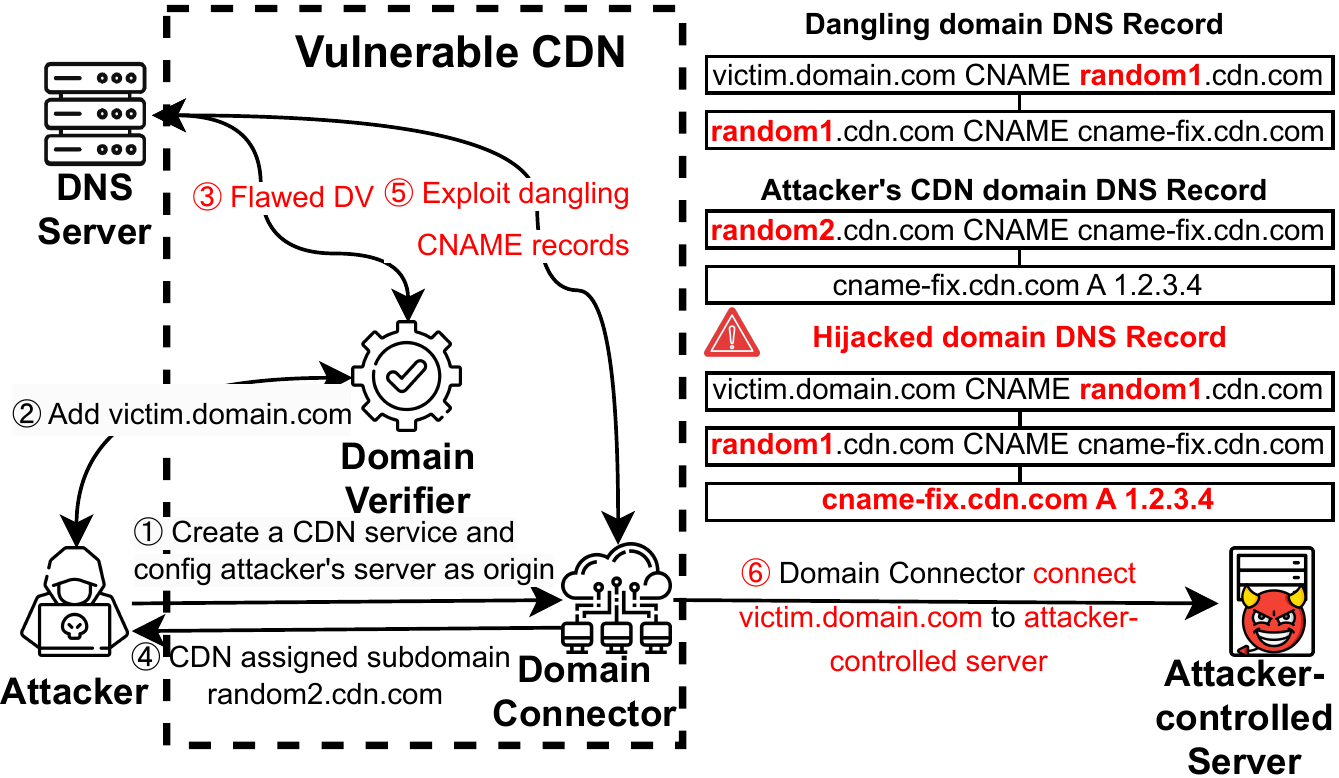}\label{fig: break_cname1}}\hspace{0.3cm}
    \subfigure[Domain takeover by exploiting CDN assigning subdomain vulnerability.]{\includegraphics[width=0.488\textwidth]{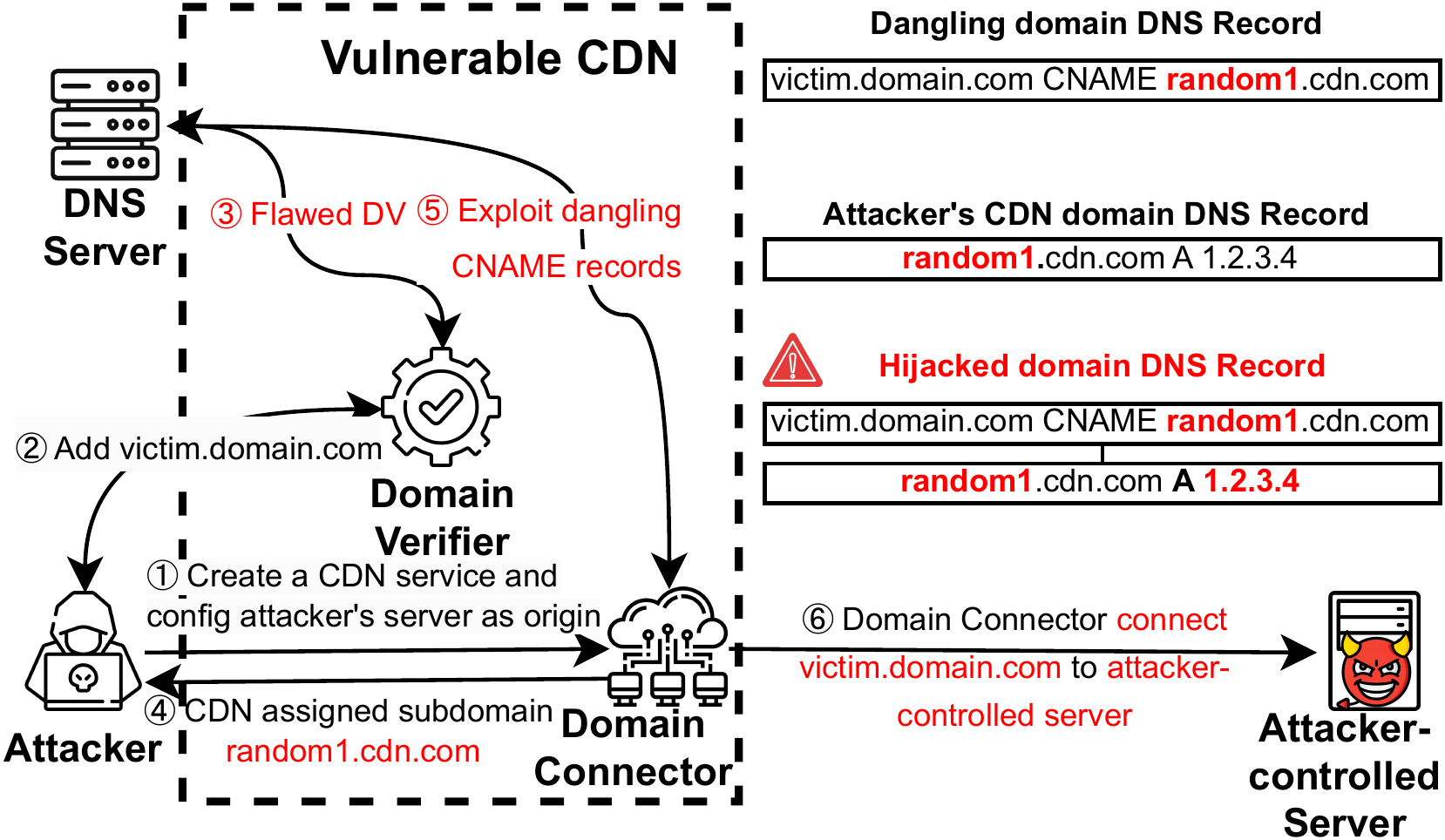}\label{fig: break_cname2}}
    \caption{Two new ways of domain takeover.}
    \label{fig: break cname}
\end{figure*}

\noindent \textbf{Finding 1: Two new ways of domain takeover.} 
When analyzing the CDN domain verification mechanism, we discover two new vulnerabilities. 
Firstly, we identify that two CDN providers (Edgio and Cachefly) have vulnerabilities in their domain connections. This vulnerability allows attackers to exploit them and bypass the CDN domain verification mechanism, leading to domain takeover. Secondly, we discover that two CDN providers (Kuocai and Yundun) employ domain-based random generation of CDN-assigned subdomains. However, this method is flawed, as attackers can generate the same CDN-assigned subdomain, which could result in domain takeover.
These vulnerabilities suggest that the CDN domain verification mechanism may not sufficiently consider the potential threat of attackers, leaving the CDN domain verification mechanism vulnerable.

W1: Domain takeover by exploiting CDN domain misconnection. As shown in Figure~\ref{fig: break_cname1}, the dangling domain's DNS record points to the CDN-assigned subdomain (i.e., random1.cdn.com) is random and unique. Then, the CDN further resolves it to a fixed subdomain (i.e., cname-fix.cdn.com). Attackers can create a CDN service and add ``victim.domain.com'' as the custom domain (\ding{192} \ding{193} \ding{194}). The CDN then assigns a different subdomain (i.e., random2.cdn.com) (\ding{195}). However, due to misconnection, the CDN connects ``victim.domain.com'' to the attacker-controlled server via the fixed subdomain (\ding{196} \ding{197}). 

W2: Domain takeover by exploiting CDN assigning subdomain vulnerability. We also found another vulnerability in the Domain Connector where the CDN assigned the same random subdomain when different accounts added the same custom domain. As shown in Figure~\ref{fig: break_cname2}, the dangling domain's DNS record points to the CDN-assigned subdomain (i.e., random1.cdn.com) is random and unique. 
However, attackers can create a CDN service and add ``victim.domain.com'' as the custom domain to obtain the same CDN-assigned subdomain (i.e., random1.cdn.com) (\ding{192} \ding{193} \ding{194} \ding{195}). Then, attackers exploit dangling domains, and CDN connects the victim.domain.com to the attacker-controlled server (\ding{196} \ding{197}).  

\noindent \textbf{Finding 2: Exploiting Multi-CDN to domain takeover.}
Due to the integration of multiple CDNs in Multi-CDN, users can choose one of the CDN providers to provide services for them. 
However, it also allows attackers to exploit the vulnerable Multi-CDN to take over the dangling domains hosted by secure CDNs. We discover that KuaikuaiCloud conducts domain verification based on the custom domain. If the domain is verified, another account (attacker) can claim to own the custom domain. KuaikuaiCloud generates the same CDN-assigned subdomain as Baidu based on the customized domains (i.e., custom.com.a.bdydns.com). Therefore, attackers can bypass Baidu's domain verification mechanism by creating services on KuaikuaiCloud and adding these dangling domains as custom domains to take over these dangling domains. However, it requires the domains initially hosted by KuaikuaiCloud, then moved to Baidu. Eventually, the customer terminated the CDN service, which resulted in dangling domains hosted by a secure CDN. 

\noindent \textbf{Finding 3: Exploiting Multi-CDN to domain fronting.} 
We find that some CDN providers adopt cloud fusion technology, enabling them to leverage the infrastructure of multiple CDN. 
By subscribing to a Multi-CDN service for a domain, it is possible to dynamically associate with different CDNs based on metrics such as latency, performance overhead, proximity, and other factors. 
While researching the domain fronting, we discover two fascinating phenomena. Firstly, attackers can create a Multi-CDN service and set up a high-reputation domain hosted by another CDN as SNI to launch the domain fronting. For example, KuaikuaiCloud integrates Baidu and Tencent. Customers can choose one to provide services when they deploy their domain on KuaikuaiCloud. Among these two CDN providers, Baidu is vulnerable to domain fronting. 
Therefore, attackers can exploit KuaikuaiCloud to abuse high-reputation domains hosted on Baidu for domain fronting, reducing the risk of being blocked. 
Secondly, we know CloudFront banned domain fronting in 2018~\cite{dfban}. However, AligeCDN uses CloudFront's infrastructure and can use CloudFront's shared certificates, which resurrected domain fronting in CloudFront. Attackers can exploit AligeCDN to launch domain fronting and use CloudFront's share certificates to circumvent network censorship.

\section{DISCUSSION}

\subsection{Ethical Considerations}
We consciously designed our automatic detection system to avoid raising network alerts or causing harm to the analyzed targets. 
First, the system measurement phase has been carried out with simple DNS queries and benign HTTP requests. 
Second, we distribute HTTP requests to different CDN ingress nodes at a low rate, which is negligible compared with CDN's global service capability.
Third, we only identify vulnerable CDN providers and domains and do not exploit these domains for any malicious activity. 
Last, we did not receive complaints or requests from the analyzed CDN providers and website owners to opt out of future scans. 
As a result, we are confident that our measurements had no measurable impact on either the CDN infrastructure or the websites behind the CDNs.
In addition, we would like to emphasize that our open-source measurement system provides a secure and reliable solution for security professionals to assess their company websites for vulnerabilities. Individuals can use our system to measure their websites, identify and address vulnerabilities, and even opt for a more secure CDN. It is important to note that users can confidently utilize our system without affecting the performance or availability of the CDNs or the websites behind the CDNs.

\subsection{Responsible Disclosure}
We have already disclosed our findings to all vulnerable CDNs. \lzy{Up until now, twelve CDN providers have confirmed the vulnerabilities. Edgio, Kuocai, Yundun, and KuaikuaiCloud have acknowledged the new attack surface we discovered and are working to fix it. ChinaNetCenter has acknowledged the vulnerability and is working to find solutions. Fastly and G-core fixed the vulnerability after our disclosure.}
Baidu, Bunny, Cachefly, and UCloud acknowledged the issue but argued that as CDN providers, ``we provide services to our customers and are not responsible for how they configure them''. Netlify acknowledged the vulnerability and decided there were no direct security implications for their platforms through the HackerOne platform. The other CDN providers requested vulnerability details, and we are still awaiting further responses.

\subsection{Mitigation}
DVA vulnerabilities have been known for quite some time. However, CDN providers have not yet fully addressed these vulnerabilities. Therefore, it is of utmost importance for us to actively carry out measurements and report any identified vulnerabilities to CDN providers. Moreover, website owners should consider using our system to monitor their domains hosted on CDNs. They should also take prompt action upon discovering any vulnerabilities. By taking these proactive steps, we can collectively enhance the security and resilience of CDN infrastructure and protect against potential attacks.

\noindent \textbf{Domain fronting mitigation.} 
Akamai, a well-known CDN provider, offers an example of how a CDN typically handles incoming HTTPS requests~\cite{request_flow}. The client first establishes a TLS session with the CDN and sends the TLS certificate to the CDN to validate that the session is secure. The client then sends an HTTPS request to the CDN, which forwards the request to the origin server. To mitigate domain fronting, we recommend CDN verify that the Host header in all HTTPS requests matches the SNI in the TLS session.

\noindent \textbf{Domain borrowing mitigation.} 
We have mitigations from the perspectives of CDN providers and website owners. CDN providers must perform strict domain verification using standardized procedure~\cite{dov}. For example, providers can set up random values in TXT records~\cite{TXT_verification}, one-time tokens, or TLS certificates. When comparing TLS certificate verification~\cite{aas2019let} to DNS record verification, it is evident that TXT record verification is more effective. Attackers can obtain TLS certificates illicitly, such as by buying a domain, applying for its TLS certificate, and refunding the domain. Website owners should utilize the certificate transparency mechanism~\cite{laurie2014certificate} to promptly detect when attackers apply new TLS certificates for their domains. If attackers steal their TLS certificate, the administrators should as soon as possible revoke the certificate to prevent further abuse~\cite{smith2020let}.

\noindent \textbf{Domain takeover mitigation.} Based on our analysis, we have concluded that the following strategies are recommended practices to prevent domain takeover. Firstly, providers must perform strict Domain Verification as described above.
Secondly, when different users deploy the same custom domain, CDN providers must avoid potential CDN-assigned subdomain collisions. 
Thirdly, CDN providers should keep records of all historical user-defined labels or hosted domains and their associated accounts to prevent attackers from easily customizing CDN-assigned subdomains identical to the victims. 
Fourthly, CDN providers should perform critical checks before the customer terminates CDN service~\cite{azbantake}. They should verify that customers delete the DNS records associated with that CDN service before termination.
\section{Conclusion}
We systematically explored the CDN domain verification mechanism to examine the existing and unveil new DVA flaws in CDN implementation. To understand the severity, we designed and implemented an automatic detection system, \toolname, to comprehensively evaluate DVA threats on the Internet. We recognized that 43 CDN providers are vulnerable to DVA threats, including popular CDN providers such as Akamai and Fastly. We also discovered two new ways of domain takeover and unveiled that CDN providers sharing infrastructures exacerbate DVA threats. We measured 89M subdomains from Tranco Top 1M domains and discovered over 20k borrowing domains and 1k dangling domains. Our findings highlighted the need for more research efforts to improve CDN security practices and reduce DVA threats. 
\section{Compliance with the Open Science Policy}
We will open the source of DVAHunter through GitHub at this link (\href{https://github.com/LinZiyuu/DVAHunter}{https://github.com/LinZiyuu/DVAHunter}).

\bibliographystyle{unsrt}
\normalem
\bibliography{reference}

\appendix

\section{Collecting exposed origin IPs.} 
The increasing prevalence of Distributed Denial of Service (DDoS) attacks on the internet has led to the widespread adoption of DDoS Protection Services \cite{gilad2016cdn,guo2023temporal,zhengreqsminer, CDN_Cannon}, which are typically provided by CDNs and integrated with the CDN's security infrastructure. The effectiveness of the CDN relies heavily on hiding the IP address of the origin server and rerouting traffic to the distributed infrastructure of the CDN providers, where malicious traffic can be blocked. In this paper, we discovered through CNAME analysis that some new CDN providers have residual resolution vulnerabilities \cite{exposure_origin_ip}. This vulnerability refers to the situation where the CDN provider directly points the domain to the origin IP address when a customer terminates their service or switches to another CDN provider, resulting in the leakage of the origin IP address. It renders the protection provided by future CDN providers ineffective, as attackers can discover the origin IP address and launch DDoS attacks directly against the origin server. We identified three major CDN providers, EdgeNext, Yundun, and Yunaq, that are vulnerable to this residual resolution exposure. Typically, these CDN providers resolve domain names to multiple ingress nodes, allowing customers to choose the nearest ingress node. However, we reveal that when website customers shut down their CDN service, three CDN providers resolve customers' websites directly to their origin IPs, which can then leak the IPs of customers' origin servers to attackers. We verified the exposure of origin IPs by obtaining domain names with only one A record hosted by these CDN providers and sending HTTP requests to both the domain name and the IP address, comparing their responses. We assessed the scale of the problem in the Tranco Top 1M domains. As shown in Table~\ref{table: ip exposure}, we identified 4766 domains hosted on vulnerable CDNs, and 727 domains (15.25\%) exposed their origin server IPs.

\begin{table}[h]
\caption{Origin IP exposure in the wild.}
\small
\centering
\begin{tabular}{lcc}
\toprule[1.5pt]
CDN Provider  & \# Exposed IP(\%\*) & \# Hosted Domain\\ 
\midrule

EdgeNext & 18 (0.6\%)            & 1958      \\
Yunaq    & 705 (25\%)          & 2748      \\
Yundun   & 4 (6.66\%)            & 60        \\ 
\midrule
All   & 727 (15.25\%)            & 4766        \\ 
\bottomrule[1.5pt]
\end{tabular}
\label{table: ip exposure}
\end{table}

\section{Benefits of Our System}
DVAHunter can detect CDNs vulnerable to DVA threats and continuously monitor which domains are hosted on vulnerable CDNs and which are at risk of domain takeover and domain borrowing. It is beneficial to a diverse group of stakeholders. Firstly, our system can help cybersecurity professionals understand how attackers exploit DVA to abuse domains hosted by CDNs. It can help them effectively monitor and detect malicious activities. Secondly, CDN customers can use our system to assess and analyze whether their business is vulnerable to DVA threats, helping them make more informed decisions when choosing CDN providers. Additionally, CDN providers can leverage our system to monitor their vulnerability to DVA threats and rectify any flaws in their domain verification mechanism. They can also conduct regular scans of the domains hosted on their CDNs to identify potential takeover risks and send email notifications to domain owners, urging them to remove dangling DNS records.

\section{Anonymity and Cost}
\label{sec:set-diff-dodis}
In this study, it is described that attackers can exploit the DVA vulnerability for malicious purposes. However, it will discourage attackers when the CDN service needs to be paid or could expose attackers' identities. 
Unfortunately, as shown in Table~\ref{table: all anonymity and cost}, CDN providers often offer ``free'' or ``free trial'' services well as do not require the customers' identifying information, which makes it easier for attackers to launch attacks from exploiting these CDN providers.
Of all the CDN providers, 21 only require a valid email address, including Cloudflare, Fastly, KeyCDN, etc.
Seven CDN vendors, including Azure, Cachefly, and CloudFront, require valid credit card information (which could be a gift card or stolen card).
Ten CDN vendors, including Alibaba, ChinaNetcenter, and Tencent, require a valid phone number (which could be an anonymous phone number). 
Six CDN vendors, including Alibaba, HuaweiCloud, and Ucloud, require users to verify their identity with a valid bank card, which makes it more challenging for attackers to remain anonymous.

\begin{table}[hb]
\centering
\caption{CDN registration requirements, cost, and domain verification.}
\begin{threeparttable}
\small
\begin{tabular}{@{}l@{\extracolsep{\fill}}ccc@{}}
\toprule
               \textbf{CDN Privider} & \textbf{Requirements} & \textbf{Price} & \textbf{DV} \\
\midrule
Akamai                & N/A                            & N/A            & N/A              \\
Alibaba               & C1, C2, C4                     & Free trial     & DNS              \\
AligeCDN                 & C1                             & Free trial     & No   \\
ArvanCloud            & C1, C3                         & Free trial     & N/A\textsuperscript{$\star$}              \\
Azion                 & C1                             & Free trial     & No   \\
Azure                 & C1, C3                         & Paid           & No   \\
Baidu                 & C1                             & Free trial     & DNS              \\
BelugaCDN             & C1, C2, C3                     & Paid           & N/A\textsuperscript{$\star$}              \\
Bunny                 & C1                             & Free trial     & No   \\
Cachefly              & C1, C3                         & Free service   & No   \\
CDN77                 & C1                             & Free trial     & No   \\
CDNetworks            & C1                             & Free trial     & No   \\
CDNsun                & C1                             & Free trial     & No   \\
CDNvideo              & C1, C2                         & Paid           & N/A\textsuperscript{$\star$}              \\
ChinaNetCenter        & C1, C4                         & Free trial     & No   \\
Cloudflare            & C1                             & Free service   & DNS              \\
CloudFront            & C1, C3                         & Free service   & TLS Certificate \\
DogeCloud             & C1, C2                         & Free trial     & DNS              \\
EdgeNext              & C1                             & Free trial     & No   \\
Edgio                 & C1                             & Free trial     & No   \\
Fastly                & C1                             & Free service   & No   \\
G-core                & C1                             & Free service   & DNS              \\
HuaweiCloud                & C1, C4                         & Paid           & DNS              \\
JDCloud               & C1, C4                         & Paid           & DNS              \\
KeyCDN                & C1                             & Free trial     & No   \\
KingsoftCloud         & C1, C4                         & Paid           & DNS              \\
KuoCai                & C1                             & Free trial     & DNS              \\
KuaikuaiCloud         & C1                             & Free trial     & DNS              \\
LeaseWeb              & C1, C2                         & Paid           & N/A\textsuperscript{$\star$}              \\
Layun                 & C1                             & Free trial     & DNS              \\
LightCDN              & C1                             & Free trial     & No   \\
Limelight\textsuperscript{$\wr$}             & N/A                            & N/A            & N/A              \\
Lumen\textsuperscript{$\wr$}                 & N/A                            & N/A            & N/A              \\
Medianova             & C1                             & Free trial     & No   \\
Netlify               & C1, C3                         & Free service   & No   \\
Qiniu                 & C1, C2                         & Free service   & DNS              \\
StackPath             & C1, C3                         & Free trial     & No   \\
Tencent               & C1, C2                         & Free trial     & DNS              \\
UCloud                & C1, C2, C4                     & Paid           & No   \\
Udomain               & C1                             & Free trial     & No   \\
UPYun                 & C1, C2                         & Free trial     & DNS              \\
Yunaq                 & C1, C2                         & Free trial     & DNS              \\
Yundun                & C1                             & Free trial     & DNS              \\
Sudun                 & C1                             & Free trial     & DNS               \\
\bottomrule 
\end{tabular}
\begin{tablenotes}

\item[$\dagger$] C1 means an Email address is required to register an account.
\item[$\ddagger$] C2 means a Phone number is required to register an account.
\item[$\rVert$] C3 means a Credit card is required to register an account.
\item[$\ast$] C4 means a Bank card is required to register an account.
\item[$\star$] We failed to register an account.
\item[$\wr$] Limelight has been acquired by Edgio, while Lumen has been acquired by Akamai.
\end{tablenotes}
\end{threeparttable}
\label{table: all anonymity and cost}
\end{table}

\section{Summary of vulnerable CDN providers.}
We have summarized the vulnerable CDN providers in Table~\ref{table: vulnerable_cdn}.
\begin{table*}[ht]
    \caption{Vulnerable CDN providers.}
    \centering
    \small
    \begin{threeparttable}
    \begin{tabular}{lccc|lccc}
    
    \toprule
               & fronting & borrowing & takeover &               & fronting      & borrowing     & takeover      \\
    \midrule
Akamai         & \XSolidBrush               & N/A              & N/A             & Huawei        & \Checkmark                    & \XSolidBrush                    & \XSolidBrush                    \\
Alibaba        & \Checkmark               & \XSolidBrush                & \XSolidBrush               & JDCloud       & \Checkmark                    & \XSolidBrush                    & \XSolidBrush                    \\
AligeCDN       & \Checkmark\textsuperscript{$\ast$}               & \XSolidBrush                & \XSolidBrush               & KeyCDN        & \Checkmark                    & \Checkmark                    & N/A                  \\
ArvanCloud     & \Checkmark               & N/A              & N/A             & Kingsoft      & \Checkmark                    & \XSolidBrush                    & \XSolidBrush                    \\
Azion          & \Checkmark               & \Checkmark                & \XSolidBrush               & KuaikuaiCloud & \Checkmark                    & \XSolidBrush                    & \Checkmark\textsuperscript{$\rVert$}                    \\
Azure          & \XSolidBrush               & \XSolidBrush                & \Checkmark               & Kuocai        & \Checkmark                    & \Checkmark                    & \Checkmark \textsuperscript{$\ddagger$}                   \\
Baidu          & \Checkmark               & \XSolidBrush                & \XSolidBrush               & Layun         & \Checkmark                    & \Checkmark                    & \Checkmark                    \\
Belugacdn      & \Checkmark               & N/A              & N/A             & LeaseWeb      & \Checkmark                    & N/A                  & N/A                  \\
Bunny          & \Checkmark               & \Checkmark                & \Checkmark               & LightCDN      & \Checkmark                    & \Checkmark                    & \Checkmark                    \\
Cachefly       & \Checkmark               & \Checkmark                & \Checkmark\textsuperscript{$\dagger$} & Limelight     & \Checkmark                    & N/A                  & N/A                  \\
CDN77          & \Checkmark               & \Checkmark                & \XSolidBrush               & Lumen         & \Checkmark                    & N/A                  & N/A                  \\
CDNetworks     & \Checkmark               & \Checkmark                & \Checkmark               & Medianova     & \Checkmark                    & \Checkmark                    & \XSolidBrush                    \\
CDNSun         & \Checkmark               & N/A              & N/A             & Netlify       & \Checkmark                    & \Checkmark                    & \XSolidBrush                    \\
CDNvideo       & \XSolidBrush               & \XSolidBrush                & \XSolidBrush               & Qiniu         & \Checkmark                    & \XSolidBrush                    & \XSolidBrush                    \\
ChinaNetCenter & \Checkmark               & \Checkmark                & \Checkmark               & StackPath     & \Checkmark                    & \Checkmark                    & \XSolidBrush                    \\
Cloudflare     & \XSolidBrush               & \XSolidBrush                & \Checkmark              & Sudun         & \Checkmark                    & \Checkmark                    & \Checkmark                    \\
CloudFront     & \XSolidBrush               & \XSolidBrush                & \XSolidBrush               & Tencent       & \XSolidBrush                    & \XSolidBrush                    & \XSolidBrush                    \\
DogeCloud      & \Checkmark               & \Checkmark                & \Checkmark               & UCloud        & \Checkmark                    & \Checkmark                    & \Checkmark                    \\
EdgeNext      & \Checkmark               & \Checkmark                & \Checkmark               & Udomain       & \Checkmark                    & \Checkmark                    & \XSolidBrush                    \\
Edgio          & \Checkmark               & \Checkmark                & \Checkmark\textsuperscript{$\dagger$} & UPYun         & \Checkmark                    & \XSolidBrush                    & \XSolidBrush                    \\
Fastly         & \Checkmark               & \Checkmark                & \Checkmark               & Yunaq         & \Checkmark                    & \XSolidBrush                    & \XSolidBrush                    \\
G-core         & \Checkmark               & \XSolidBrush                & \Checkmark           & Yundun        & \Checkmark                    & \Checkmark                    & \Checkmark\textsuperscript{$\ddagger$}                    \\
Goooood        & \Checkmark               & \Checkmark                & \XSolidBrush               &               & \multicolumn{1}{l}{} & \multicolumn{1}{l}{} & \multicolumn{1}{l}{}\\
\bottomrule
\end{tabular}

   \begin{tablenotes}
    \item[$\dagger$] We discover a new attack surface in this CDN provider regarding domain takeover (W1).
    \item[$\ddagger$] We discover a new attack surface in this CDN provider regarding domain takeover (W2).
    \item[$\rVert$] We discover that an attacker could use this Multi-CDN to take over a dangling domain hosted by a secure CDN.
    \item[$\ast$] N/A means that the CDN provider is for businesses only or does not offer a free trial.
    \end{tablenotes}
    \end{threeparttable}    
    
    \label{table: vulnerable_cdn}
\end{table*}

\end{document}